\documentclass[twocolumn,twocolappendix]{aastex6}

\usepackage{amsmath}

\begin{document}

\title{Bayesian Estimation of Thermonuclear Reaction Rates}

\author{
C.~Iliadis,\altaffilmark{1,2}
K.S.~Anderson,\altaffilmark{1}
A.~Coc,\altaffilmark{3}
F.X.~Timmes\altaffilmark{4,5}
S.~Starrfield\altaffilmark{4}
}

\altaffiltext{1}{Department of Physics \& Astronomy, University of North Carolina at Chapel Hill, Chapel Hill, NC 27599-3255, USA}
\altaffiltext{2}{Triangle Universities Nuclear Laboratory, Durham, NC 27708-0308, USA}
\altaffiltext{3}{Centre de Sciences Nucl\'eaires et de Sciences de la Mati\`ere (CSNSM), CNRS/IN2P3, Univ. Paris-Sud, Universit\'e Paris--Saclay, B\^{a}timent 104, F-91405 Orsay Campus, France}
\altaffiltext{4}{School of Earth and Space Exploration, Arizona State University, Tempe, AZ 85287-1504, USA}
\altaffiltext{5}{Joint Institute for Nuclear Astrophysics}

\email{iliadis@unc.edu}

\begin{abstract}
The problem of estimating non-resonant astrophysical S-factors and thermonuclear reaction rates, based on measured nuclear cross sections, is of major interest for nuclear energy generation, neutrino physics, and element synthesis. Many different methods have been applied in the past to this problem, almost all of them based
on traditional statistics. Bayesian methods, on the other hand, are now in widespread use in the physical sciences. In astronomy, for example, Bayesian statistics is applied to the observation of extra-solar planets, gravitational waves, and type Ia supernovae. However, nuclear physics, in particular, has been slow to adopt Bayesian methods. We present astrophysical S-factors and reaction rates based on Bayesian statistics. We develop a framework that incorporates robust parameter estimation, systematic effects, and non-Gaussian uncertainties in a consistent manner. The method is applied to the d(p,$\gamma$)$^3$He, $^3$He($^3$He,2p)$^4$He, and $^3$He($\alpha$,$\gamma$)$^7$Be reactions, important for deuterium burning, solar neutrinos, and big bang nucleosynthesis.
\end{abstract}

\keywords{methods: numerical --- nuclear reactions, nucleosynthesis, abundances --- stars: interiors --- primordial nucleosynthesis}

\section{Introduction}\label{sec:intro}
Thermonuclear reaction rates are at the heart of nuclear astrophysics. They are essential for understanding key phenomena in the universe, including main-sequence stars, red giants, AGB stars, white dwarfs, core-collapse and thermonuclear supernovae, classical novae, and type I X-ray bursts. The evaluations provided by Willy Fowler and collaborators were of outstanding importance in this regard \citep[][]{fowler67,fowler75,caughlan88}. Their recommended {\it experimental} reaction rates provided, for the first time, a solid nuclear physics foundation for models of stars and big bang nucleosynthesis. A further milestone was reached with the NACRE collaboration \citep[][]{angulo99}. Their work provided not only updated rates, but also included approximate reaction rate error estimates. These ideas were subsequently extended to heavier target nuclei and to reactions involving short-lived targets \citep[][]{iliadis01}. 

In recent years, a growing volume of astronomical data motivated an increased number of nucleosynthesis sensitivity studies to better quantify the impact of given reactions on nuclear burning. The first such reaction network studies utilized published recommended thermonuclear reaction rates together with somewhat arbitrary methods of varying the rates \citep[][]{iliadis02,stoesz03,rapp06,parikh08,iliadis11}. It became apparent that improved experimental reaction rate estimates, based on sound statistical methods, would be very valuable. Such experimental rates were first published in 2010 \citep[][]{longland10,iliadis10a,iliadis10b,iliadis10c}. They were obtained using Monte Carlo sampling of the many experimental nuclear physics input quantities (e.g., resonance energies and strengths, partial widths and reduced widths) entering in a reaction rate calculation. 

The output of this procedure is a probability density for the reaction rate at each temperature of interest. The probability density is used to extract statistically meaningful rate estimates, such as a recommended rate (from the median) or rate uncertainties (from the 16th and 84th percentiles for a 68\% coverage probability). Experimental Monte Carlo-based reaction rates are tabulated in the STARLIB reaction rate library \citep[][]{sallaska13} and are publicly available.\footnote{\url{http://starlib.physics.unc.edu/index.html}} The STARLIB library has already been used in Monte Carlo nucleosynthesis studies of classical novae \citep[][]{kelly13} and in studies of globular cluster polluters \citep[][]{iliadis16}. Recently, STARLIB has been used\footnote{\url{https://github.com/carlnotsagan/ReacSamp}} with the MESA stellar evolution software instrument \citep[][]{paxton11,paxton13,paxton15} to study the impact of uncertainties in nuclear reaction rates on the properties of carbon-oxygen white dwarfs \citep[][]{fields16}.

Experimental Monte Carlo-based thermonuclear reaction rates are so far available for $65$ (p,$\gamma$), (p,$\alpha$), and ($\alpha$,$\gamma$) reactions in the $A$ $=$ $14-40$ mass region, involving both stable and unstable target nuclei. The Monte Carlo-based method of estimating reaction rates is limited, in its present form, to nuclear reactions that are dominated by resonant contributions to the total rate. Non-resonant contributions are included in the method \citep[][]{longland10}, but their random sampling is only performed in the simplest possible manner by providing an approximate uncertainty of the non-resonant astrophysical S-factor. While this treatment of the non-resonant component is not statistically rigorous, it has little practical effect on the total rates for the reactions referred to above, precisely because they are dominated by resonant contributions.

The calculation of non-resonant reaction rates\footnote{With the expression ``non-resonant'', we refer to astrophysical S-factors that vary smoothly with energy.} directly from experimental data has its own difficulties and pitfalls. Such rates have been estimated for $10$ light-particle reactions, in the $A$ $=$ $2-7$ mass region, using the R-matrix reaction model and $\chi^2$ fits to the data for the purpose of studying big bang nucleosynthesis \citep[][]{descouvemont04}. Light-particle reaction rates, in the $A$ $=$ $2-18$ mass range, have also been computed for solar models \citep[][]{adelberger11}. The experimental data were analyzed in the latter work by $\chi^2$ minimization, using either a polynomial S-factor expansion or the output of theoretical nuclear reaction models. Typical problems encountered in the analysis of non-resonant rates include the treatments of data normalization factor uncertainties (i.e., systematic errors) and discrepant data sets. In a recent study of the cosmic evolution of deuterium \citep[][]{coc15}, a number of different methods, all based on $\chi^2$ minimization, have been employed to compute rates for the d(p,$\gamma$)$^{3}$He, d(d,n)$^{3}$He, and d(d,p)$^3$H reactions.  

In this work we provide a fresh look by calculating the non-resonant reaction rates using Bayesian probability theory. The advantages of this approach are manifold. First, the Bayesian approach yields directly the quantity of interest in nucleosynthesis sensitivity studies, i.e., the reaction rate probability density function. These rates can be easily implemented, together with the Monte Carlo-based rates discussed above, into the STARLIB rate library. Second, the Bayesian model provides a more consistent method for extracting information from measured data, even in ill-conditioned situations, compared to traditional statistics.

In Section~\ref{sec:regression2}, we will discuss how to incorporate systematic uncertainties, robust regression, and non-Gaussian statistical uncertainties into a Bayesian analysis. Bayesian astrophysical S-factors and thermonuclear rates for the d(p,$\gamma$)$^3$He, $^3$He($^3$He,2p)$^4$He, and $^3$He($\alpha$,$\gamma$)$^7$Be reactions are presented in Sections~\ref{sec:sfactors} and  \ref{sec:rates}, respectively.  A summary and conclusions are provided in Section \ref{sec:summary}. Since Bayesian inference has rarely been applied before to S-factors and reaction rates\footnote{For an interesting application of Bayesian methods to estimate the parameters of effective field theories, and an application to the S-factor and reaction rates of $^7$Be(p,$\gamma$)$^8$B, see \citet[][]{Zha15,Wes16}.}, we will discuss in Appendix~\ref{sec:inference} how to use Bayes' theorem to estimate model parameters. To clarify our discussion of Bayesian S-factors and reaction rates in the main text, these ideas are applied in Appendix~\ref{sec:regression} to the simple problem of linear regression.

\section{Systematic uncertainties, robust regression, and non-Gaussian statistical uncertainties}\label{sec:regression2}
For the analysis of Bayesian models, we will employ the program \texttt{JAGS} (``Just Another Gibbs Sampler'') using Markov chain Monte Carlo (MCMC) sampling. More information on \texttt{JAGS}, including a simple example, is provided in the Appendices. Before we can analyze astrophysical S-factor data using Bayesian inference, we have to consider how to include systematic uncertainties, outliers, and non-Gaussian statistical uncertainties in the likelihood function.

\subsection{Systematic uncertainties}\label{sec:sysuncert}
Experimental data are subject to statistical and systematic uncertainties. Statistical uncertainties are well understood and they usually follow a known probability distribution, e.g., a Gaussian or Poissonian. When a series of independent experiments is performed, statistical uncertainties will give rise to different results in each individual measurement. The magnitude of the statistical uncertainty can be estimated from the standard deviation of the data, if the experiments are uncorrelated. Statistical effects can be reduced by combining the results from several measurements.

Systematic effects, on the other hand, do not usually signal their existence by a larger fluctuation of the data. When the experiment is repeated, the presence of systematic effects may not produce different answers. Similarly, systematic uncertainties are frequently not reduced when combining the results from different measurements. Reported systematic uncertainties are at least partially based on assumptions made by the experimenter, are model dependent, and follow vaguely known probability distributions \citep[][]{heinrich07}. 

Consider as an example the measurement of an astrophysical S-factor at a given bombarding energy. The experimental result is frequently reported as $S_{mean}$ $\pm$ $\sigma_{stat}$ $\pm$ $\xi_{sys}$, where $S_{mean}$ is the mean value, $\sigma_{stat}$ is the standard deviation representing the statistical uncertainty, and $\xi_{sys}$ denotes the systematic uncertainty. The latter two quantities are either reported as absolute or relative (i.e., percent) uncertainties. If a single uncertainty is required, statistical and systematic uncertainties can be combined in quadrature on the grounds that they are uncorrelated. Since a systematic effect will shift all points of a given data set in the same direction, it can either be quantified as an (additive) offset or a (multiplicative) normalization. The true value of the offset or normalization is, of course, unknown, otherwise there would be no systematic uncertainty. However, we do have one piece of information: the expectation value of the systematic uncertainty is zero, if the systematic effect is quantified as an offset, or unity, if it is described as a normalization. If this would not be the case, we would have corrected the data for the systematic effect.

The S-factor data we will analyze in Section~\ref{sec:sfactors} have been reported with systematic uncertainties described by normalization factors. For example, suppose the systematic uncertainty for a given data set is reported as $\pm10\%$, implying that the normalization uncertainty is given by a factor of $1.10$. A useful distribution for factor uncertainties is the  lognormal probability density, given by
\begin{equation}
f(x) = \frac{1}{\sigma \sqrt{2\pi}x}e^{-(\ln x - \mu)^2/(2\sigma^2)},\,\,x>0
\label{eq:lognormal}
\end{equation}
It is characterized by two quantities, the location parameter, $\mu$, and the spread parameter, $\sigma$. Notice that $\mu$ and $\sigma$ are not the mean and standard deviation of the lognormal distribution, but of the Gaussian distribution for $\ln x$. The median value of the lognormal distribution is given by $x_{med}$ $=$ $e^\mu$, while the factor uncertainty, for a coverage probability of 68\%, is $f.u.$ $=$ $e^\sigma$. Therefore, we include in our Bayesian model a systematic effect as a highly-informative, lognormal prior with a median of $1.0$, i.e., $\mu$ $=$ $0$, and a factor uncertainty given by the systematic uncertainty, i.e., in the above example, $f.u.$ $=$ $1.10$ or $\sigma$ $=$ $\ln (1.10)$. A specific example, including the syntax, for implementing systematic uncertainties into \texttt{JAGS} is given in Appendix~\ref{sec:regression}.

\subsection{Robust regression}
Outliers can bias the data analysis significantly if they are not properly taken into account. The frequently applied procedure of disregarding data points that are subjectively deemed to be outliers has no statistical justification. A number of different approaches have been applied in Bayesian inference to include outliers in the analysis. For example, the data could be described by applying a distribution that has taller tails, such as the $t$ distribution \citep[][]{lange89}, compared to the ubiquitous Gaussian distribution. 

In the present work, we adopt a different approach that is based on \citet[][]{andreon15}. The method treats the complete data set as a mixture of two populations: one population of supposedly correctly measured uncertainties, and another one for which the reported uncertainty estimates are too optimistic. The goal is to design an algorithm that can automatically identify and reduce the weight of the data points with over-optimistic uncertainties (i.e., outliers). This is achieved by including a parameter describing the membership to the different populations into the random sampling of the posterior.

The procedure has a number of advantages. In the analysis, each datum contributes to the posterior with a larger weight the smaller the uncertainty and the higher the probability that the reported uncertainty is correct. All of the data points are taken into account in the analysis and none are discarded. The MCMC sampling also quantifies the outlier probability for a given datum. The \texttt{JAGS} implementation of robust regression for a simple example is described in Appendix~\ref{sec:regression}.

\subsection{Non-Gaussian statistical uncertainties of data points}\label{sec:nongauss}
Nuclear reaction cross sections, or astrophysical S-factors, are experimentally determined by products and ratios of many nuclear physics input quantities: measured net intensities, incident beam charge, detection efficiencies, number of target nuclei, stopping powers, etc. According to the central limit theorem, the probability density of a derived quantity, such as the cross section or S-factor, will then be distributed according to a lognormal rather than a normal probability density (Equation~\ref{eq:lognormal}). This situation was discussed in \citet[][]{longland10}. The lognormal parameters are given by
\begin{equation}\label{eqn:lognorm1}
\mu =  \ln(E[x]) - \frac{1}{2}\ln\left(1 + \frac{V[x]}{E[x]^2} \right) \\
\end{equation}
\begin{equation}\label{eqn:lognorm2}
\sigma =  \sqrt{\ln \left(1 + \frac{V[x]}{E[x]^2} \right)}
\end{equation}
where $E[x]$ and $V[x]$ denote the mean value and the variance (i.e., the square root of the standard deviation), respectively. For standard deviations $\lesssim$10\% of the mean value, the lognormal probability density is very close in shape to a Gaussian. However, with increasing relative standard deviations, the differences between the lognormal density function and a Gaussian approximation increase. Notice also that, unlike the lognormal probability density, a Gaussian density function predicts a finite probability for negative values of the random variable, which is unphysical for manifestly positive quantities, such as nuclear reaction cross sections or astrophysical S-factors. The \texttt{JAGS} syntax for implementing lognormal likelihoods is given in Appendix~\ref{sec:regression}.

At low bombarding energies, where the experimental yields are very small, the reported uncertainties on data points are frequently large. For example, how should one interpret a reported S-factor of ``$30\pm15$~keVb'', which implies a finite chance of a zero S-factor? It is certainly inappropriate to assume a Gaussian likelihood in this case, because the S-factor cannot become negative. But it is equally inappropriate to assume a lognormal likelihood, which predicts zero probability for a zero S-factor. In such cases, the total statistical uncertainty is dominated by counting statistics and the appropriate likelihood function to use is a Poissonian (or a difference of Poissonians, if the net intensity is inferred from total and background counts). 

\begin{figure}[!htb]
\centering{\includegraphics[width=\columnwidth]{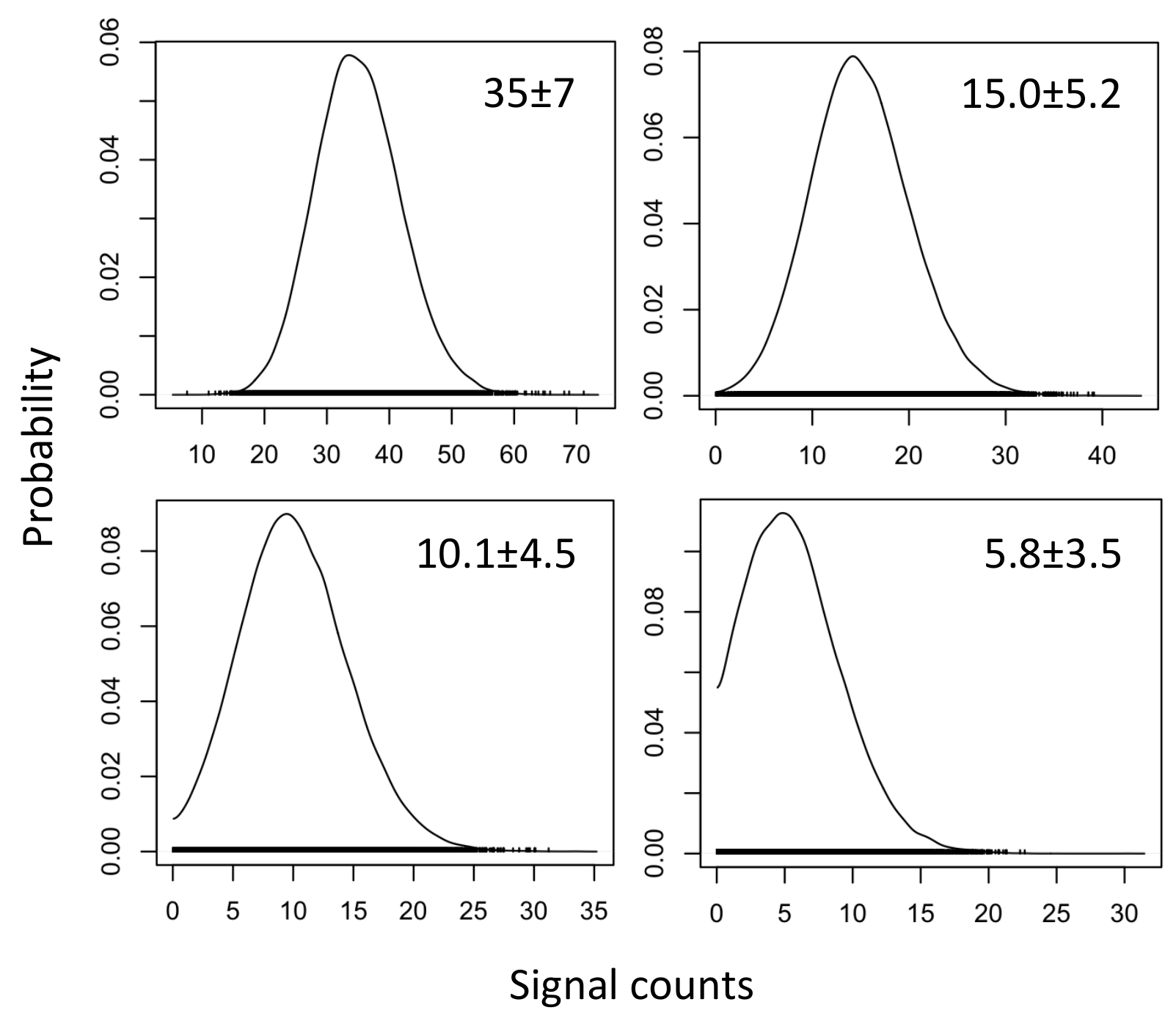}}
\caption{
Posteriors of the signal (i.e., total minus background) counts, computed using \texttt{JAGS}, for a hypothetical counting experiment. The panels are obtained for a total number of counts of  $40$ (upper left), $20$ (upper right), $15$ (lower left), and $10$ (lower right), while the background counts are kept fixed at $5$. The predicted mean and standard deviation of the signal is indicated in each panel. A uniform distribution between $0$ and $1000$ is assumed for the prior.
}\label{fig:counts}
\end{figure}

Suppose we perform a simple counting measurement. We have measured the total and the background counts, and we are interested in estimating the signal (i.e., total minus background) counts. When we set up a Bayesian model for this situation, it is appropriate to assume Poissonian likelihoods for both the total and the background counts. Figure~\ref{fig:counts} shows numerical results obtained using \texttt{JAGS}. The panels display the posteriors for the signal counts. We assumed a uniform prior between $0$ and $1000$, i.e., the posterior will closely reflect the shape of the likelihood (see Equation~\ref{eqn:bayes3}). The different panels are obtained for a total number of counts of  $40$, $20$, $15$, and $10$, while the background counts are kept fixed at $5$. The predicted mean and standard deviation of the signal is indicated in each panel. When the total number of counts is relatively large (e.g., $40$; see first panel), the probability for predicting zero signal counts is negligible. In addition, the shape of the density function is well approximated by a lognormal distribution. On the other hand, when the total number of counts is similar to the background counts (e.g., $5$; see last panel), the posterior predicts a large probability at zero signal counts and certainly does not resemble a lognormal distribution. The sequence of panels shows that the probability density can be approximated by a lognormal distribution as long as the ratio of mean value and standard deviation is $\gtrsim 3$ (see second panel). Similar results are obtained when different priors are used (e.g., gamma functions, exponentials, or hyperpriors). Consequently, we will exclude in our analysis of experimental S-factors the few data points that do not satisfy this criterion.

%Similar to the linear regression example discussed in Section~\ref{sec:regression}, we will assume that the statistical (i.e., $1\sigma$) uncertainty of each data point is described by a Gaussian likelihood. This assumption does not hold when the uncertainty is large, because in that case there is a significant probability for a negative (unphysical) value of the S-factor. Therefore, we follow the suggstion of \citep[][]{andreon10} and will only consider in our analysis data points for which the Gaussian likelihood provides an acceptable description. Specifically, we exclude data points for which the ratio of the central value and the $1\sigma$ {\it statistical} uncertainty exceeds a value of $3$. 

\section{Bayesian astrophysical S-factors}\label{sec:sfactors}
We will now apply the Bayesian method to the estimation of astrophysical S-factors\footnote{Analyzing S-factor data rather than cross section data has a number of advantages, among them a dramatic reduction in the energy dependence at the low bombarding energies considered here, and a straightforward comparison to literature results.}, $S(E)$. This quantity is defined as \citep[][]{iliadis15}
\begin{equation}
S(E) \equiv E \sigma(E) e^{2\pi\eta}\label{eq:sig2s1}
\end{equation}
where $\sigma(E)$ is the nuclear reaction cross section at the center-of-mass energy, $E$. The quantity $e^{2\pi\eta}$ denotes the Gamow factor, given by
\begin{equation}
2\pi\eta = 0.98951013 Z_0 Z_1 \sqrt{\frac{M_0 M_1}{M_0 + M_1}\frac{1}{E}}\label{eq:sig2s2}
\end{equation}
with $Z_i$ the charges of the projectile and target; in this expression, the relative atomic masses, $M_i$, and the energy, $E$, are in units of u and MeV, respectively.

The experimental S-factor can be extracted from data using fitting functions based either on a polynomial representation or on nuclear reaction models. The former provides a result that is independent of nuclear theory. Because this procedure has no theoretical justification beyond the known data points, it requires that the S-factor data cover the entire energy region of astrophysical interest. However, this is frequently not the case, especially at low bombarding energies, where the Coulomb barrier greatly inhibits direct measurements.

\begin{figure*}[!htb]
\centering{\includegraphics[width=2.0\columnwidth]{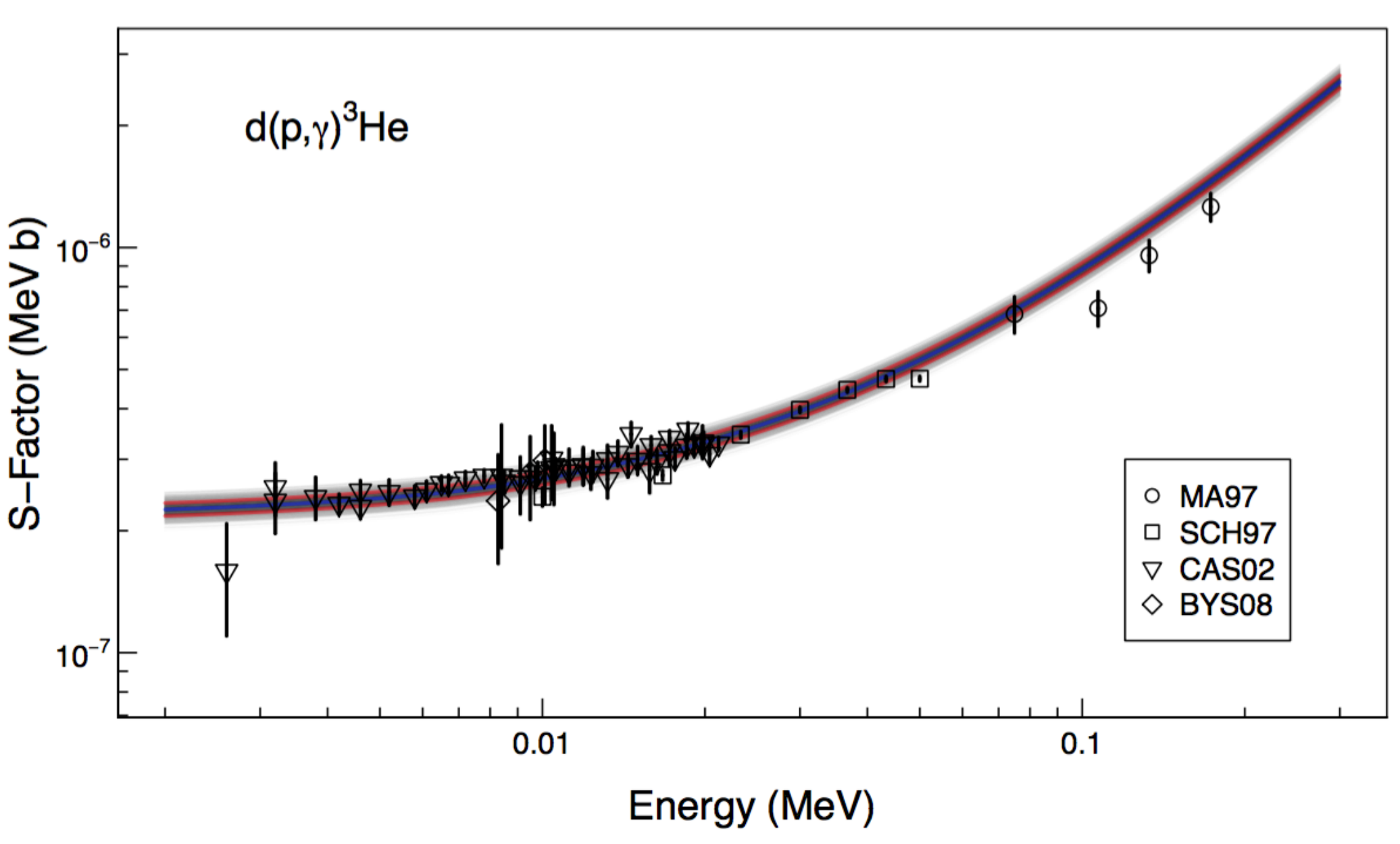}}
\caption{
Astrophysical S-factor versus center-of-mass energy for the d(p,$\gamma$)$^3$He reaction. The symbols show the data of \citet[][]{ma97} (circles); \citet[][]{schmid97} (squares); \citet[][]{casella02} (triangles); \citet[][]{bystritsky08} (diamonds). The error bars refer to ($1\sigma$) statistical uncertainties only. The lines have the following meaning: (grey shaded area) credible S-factors, obtained from the output of the \texttt{JAGS} model, where each line corresponds to one specific set of model parameters; (blue) median (50th percentile) of all credible lines; (red) 16th and 84th percentiles of all credible lines. The credible lines are calculated from the theoretical S-factor of \citet[][]{marcucci05}, multiplied by a scale factor that is a parameter of the Bayesian model. The data point at the lowest measured bombarding energy (not shown) of \citet[][]{casella02}, which has a mean-value-to-standard-deviation ratio in excess of a factor of $3$, has been omitted in our analysis (see text).
}\label{fig:dpg}
\end{figure*}

When data are missing in the region of interest, fitting functions motivated by nuclear theory (e.g., potential models, microscopic calculations, or R-matrix approaches) are usually preferred \citep[][]{adelberger11}. With this method, it is assumed that the nuclear model reliably describes the {\it energy dependence} of the S-factor, but that the {\it absolute scale} is determined by a fit of the data using the nuclear model. The assumption of an additional normalization motivated by experimental data can be explained qualitatively. For example, many microscopic models compute the interior wave functions over truncated configuration spaces, with consequences for the normalization. Similarly, in {\it ab initio} models, small variations in the strength of the effective nucleon-nucleon interaction, which is adjusted to reproduce nucleon-nucleon scattering data, will result in changes of the S-factor normalization. Nevertheless, we emphasize that the need for an additional normalization has no rigorous theoretical justification. However, since we cannot easily compute microscopic models and vary the model parameters, the assumption of a normalization factor determined by experiment represents the most straightforward method. In any case, the extrapolation of the S-factor beyond the measured data will have some theoretical justification. 

We will present in the following a Bayesian analysis for several light-particle nuclear reactions, assuming for the model S-factor either a polynomial representation or the results of nuclear models. Each of these reactions has it own intricacies.

\subsection{S-factor for d(p,$\gamma$)$^{3}$He}\label{sec:dpg}
The d(p,$\gamma$)$^{3}$He reaction represents the second step in the pp chains of stellar hydrogen burning. Since its rate is much faster compared to the first step, p(p,e$^+$$\nu$)d, uncertainties in the d(p,$\gamma$)$^3$He reaction are usually not important for stellar energy generation. In special situations, however, this reaction does play a crucial role. For example, during the earliest stages of stellar evolution, when a cloud of interstellar gas collapses to form a protostar, the central temperature reaches a few million kelvin. At this temperature, primordial deuterium fuses with hydrogen ({\it deuterium burning}), thereby generating nuclear energy that slows the contraction and the central heating of the gas until the deuterium is consumed. The d(p,$\gamma$)$^{3}$He reaction also plays a crucial role in big bang nucleosynthesis \cite[][and references therein]{coc15}, which begins when the temperature has declined to $\approx$ $0.9$~GK, corresponding to relevant kinetic energies of $\approx$ $100$~keV. The uncertainty in the d(p,$\gamma$)$^{3}$He reaction rate impacts the primordial abundances of d, $^{3}$He, and $^{7}$Li. For example, the reaction rate needs to be known to better than $\approx$ $5$\% below an energy of $200$~keV to compare big bang nucleosynthesis predictions to the very precise value (uncertainty of $1.6$\%) of the deuterium-to-hydrogen (D/H) abundance ratio measured in very metal-poor, damped Lyman-$\alpha$ systems \citep[][]{cooke14}. 

Most recently, S-factors and reaction rates for d(p,$\gamma$)$^{3}$He have been presented by \citet[][]{coc15}. A reliable estimation of S-factors requires simultaneous knowledge of statistical {\it and} systematic uncertainties, as discussed in Section~\ref{sec:sysuncert}. Among the many data sets published during 1962-2008, this information is only available for four studies \citep[][]{ma97,schmid97,casella02,bystritsky08}. These were the only data sets used by \citet[][]{coc15} for their S-factor estimation, and we will apply the same data selection\footnote{It appears that the energies in \citet[][]{bystritsky08} have been misinterpreted in \citet[][]{coc15}. The correct center-of-mass energies, used in the present work, of the three data points are $8.28$~keV, $9.49$~keV, and $10.10$~keV (instead of $8.07$~keV, $9.27$~keV, and $9.87$~keV).\label{fn:data}}. The data point at the lowest measured bombarding energy of \citet[][]{casella02} has a mean-value-to-standard-deviation ratio in excess of a factor of $3$ and has been omitted in our analysis for the reasons given in Section~\ref{sec:nongauss}. This data point was included in the analysis of \citet[][]{coc15}. The data adopted for the present analysis are displayed as black symbols in Figure~\ref{fig:dpg}, where the displayed error bars refer to ($1\sigma$) {\it statistical} uncertainties only. 
%

%%%%%%%%%%%%%%%%%%%%%%%%%%%%%%%%%%%%%%%%%%%%%%%%%%%%%%%%%%%%%%%%%%%%%%%%%%%%%%%%%%
\onecolumngrid
%\newpage
\begin{deluxetable}{cccccccc}
\tablecolumns{8}
\tablewidth{2.0\columnwidth}
\tablecaption{Results for the d(p,$\gamma$)$^3$He Reaction\label{tab:tabdpg}}
\tablehead{
\multicolumn{2}{c}{Data}  & \colhead{} & \multicolumn{2}{c}{Present\tablenotemark{a}} &   \colhead{}   & \multicolumn{2}{c}{Previous\tablenotemark{b}} \\
\cline{1-2} \cline{4-5} \cline{7-8} \\
\colhead{Ref.\tablenotemark{c}} & \colhead{n\tablenotemark{d}}   & \colhead{} & \colhead{norm\tablenotemark{e}}    & \colhead{outlier\tablenotemark{f}} & \colhead{} & \colhead{norm\tablenotemark{g}}    & \colhead{$\chi^2_\nu$\tablenotemark{h}}}
\startdata
Ma 97  & 4   & & 0.895$^{+0.058}_{-0.048}$ & 24\% & & 0.8469$\pm$0.0381 & 1.1052    \\
Sch 97 & 7   & & 0.981$^{+0.041}_{-0.041}$ & 72\% & & 0.9657$\pm$0.0062 & 11.1799  \\
Cas 02 & 51 & & 1.025$^{+0.038}_{-0.037}$ & 1.3\% & & 1.0243$\pm$0.0092 & 0.5792    \\
Bys 08 &  3  & & 1.023$^{+0.072}_{-0.068}$ & 12\% & & 1.0365$\pm$0.1457 & 0.1360 \\ 
\\
\hline
\\
\multicolumn{2}{c}{Quantity} & & \multicolumn{2}{c}{Present\tablenotemark{a}} & & \multicolumn{2}{c}{Previous\tablenotemark{b}} \\
\\
\hline
\\
\multicolumn{2}{c}{scale factor\tablenotemark{i}:} & & \multicolumn{2}{c}{$1.000^{+0.038}_{-0.036}$} & & \multicolumn{2}{c}{0.9900$\pm$0.0368} \\
\multicolumn{2}{c}{S(0) (MeVb):} & & \multicolumn{2}{c}{$(2.156^{+0.082}_{-0.077})\times10^{-7}$} & & \multicolumn{2}{c}{$(2.13\pm0.08)\times10^{-7}$} \\
\multicolumn{2}{c}{} & & \multicolumn{2}{c}{} & & \multicolumn{2}{c}{$(2.14^{+0.17}_{-0.16})\times10^{-7}$\tablenotemark{j}} \\
\multicolumn{2}{c}{} & & \multicolumn{2}{c}{} & & \multicolumn{2}{c}{$(2.1 \pm 0.4)\times10^{-7}$\tablenotemark{k}} \\
\enddata
\tablenotetext{a}{Uncertainties are derived from the 16th, 50th, and 84th percentiles of the probability density (posterior).}
\tablenotetext{b}{From \citet[][]{coc15}, unless mentioned otherwise.}
\tablenotetext{c}{Reference labels: Ma 97 \citep[][]{ma97}; Sch 97 \citep[][]{schmid97}; Cas 02 \citep[][]{casella02}; Bys 08 \citep[][]{bystritsky08}.}
\tablenotetext{d}{Number of data points in a given set.}
\tablenotetext{e}{Normalization of each data set, taking into account the reported systematic uncertainties (see text).}
\tablenotetext{f}{Probability that the data set is an outlier; computed from the average outlier probability of a given set.}
\tablenotetext{g}{Normalization of each data set, taking into account the reported systematic uncertainties (see text); the values represent $1\sigma$ uncertainties.}
\tablenotetext{h}{Reduced $\chi^2$.}
\tablenotetext{i}{Best estimate for the scale factor of the theoretical S-factor from \citet[][]{marcucci05}.}
\tablenotetext{j}{Zero-energy S-factor from \citet[][]{adelberger11}, which was obtained from a $\chi^2$ minimization of a quadratic S-factor parameterization.}
\tablenotetext{k}{Zero-energy S-factor from \citet[][]{xu13}, which was obtained using a potential model and a $\chi^2$ minimization.}
\end{deluxetable}
\twocolumngrid
%%%%%%%%%%%%%%%%%%%%%%%%%%%%%%%%%%%%%%%%%%%%%%%%%%%%%%%%%%%%%%%%%%%%%%%%%%%%%%%%%%

The S-factor fit to the data was performed in previous work by using polynomials \citep[e.g.,][]{adelberger11} or results from nuclear theory \citep[e.g.,][]{descouvemont04,coc15}. Similar to \citet[][]{coc15}, we will adopt in the present work the theoretical S-factors from \citet[][]{marcucci05}. They were obtained using variational wave functions for the p-d continuum and $^{3}$He bound states, together with a Hamiltonian consisting of two-nucleon and three-nucleon potentials. We will assume that the theoretical model adequately describes the shape of the S-factor curve, but that the absolute scale of the model S-factor is determined by the fit to the data. 

%Before discussing our Bayesian analysis, it is worth mentioning how the data were recently analyzed by \citet[][]{coc15}. For each data set from a given experiment, a normalization factor associated with the theoretical calculation of \citet[][]{marcucci05} was obtained by $\chi^2$ minimization. It turned out that the dispersion of the normalization factors was too large to justify the use of a simple weighted average. Instead, the systematic uncertainties were added quadratically to the uncertainties given by the $\chi^2$ fit. This allowed for the calculation of an average scale factor, weighted by the inflated uncertainties. 

The \texttt{JAGS} model for each of the four data sets includes the effects of systematic uncertainties, robust regression, and lognormal likelihood functions, as discussed in Section~\ref{sec:regression2}. Our Bayesian model has five parameters. Four of these are the normalizations of the individual data sets. The highly informative priors for these parameters are computed using systematic uncertainty factors of $1.09$ \citep[][]{ma97}, $1.09$ \citep[][]{schmid97}, $1.045$ \citep[][]{casella02}, and $1.08$ \citep[][]{bystritsky08}, which are listed in Table I of \citet[][]{coc15}. The fifth parameter is the common scaling factor  by which the nuclear model results have to be multiplied to fit the data. We assume a non-informative prior for this parameter, i.e., a normal probability density with a location of zero and a standard deviation of $100$. The distribution was truncated at zero since the scaling factor must be a positive quantity. Other choices of priors (i.e., uniform and gamma functions) gave very similar results. The theoretical S-factor from \citet[][]{marcucci05}, available to us as a table of 100,000 S-factor values between center-of-mass energies of $0.0$~MeV and $2.0$~MeV, was directly implemented into \texttt{JAGS}. 
%Tests showed that this procedure ensures negligible errors in the linear interpolation between grid points ($\lesssim0.007$\%).

With our Bayesian model, we generated random samples using three independent Markov chains, each of length 75,000 (without burn-in). This ensures that the Monte Carlo fluctuations are negligible compared to the statistical and systematic uncertainties. A first impression can be obtained from Figure~\ref{fig:dpg}. The grey shaded region consists of lines that correspond to the credible S-factor curves, where each line corresponds to one sampled set of model parameters. The blue line represents the median (50th percentile), and the red lines the 16th and 84th percentiles of all credible S-factors. More information on the meaning of these lines can be found in Appendix \ref{sec:inference}.

Details of our analysis are given in Table~\ref{tab:tabdpg} and are compared to recently published results obtained using traditional statistics ($\chi^2$ minimization). The top part of the table displays the normalization factors (``norm'') of each data set, taking into account the reported systematic uncertainties (see above). The present and previous values overlap within uncertainties. However, the magnitude of the uncertainties differs significantly. For example, for the data of \citet[][]{casella02} our uncertainties are a factor of $\approx$4 larger than those of \citet[][]{coc15}, while for the data of \citet[][]{bystritsky08} our uncertainties are smaller by a factor of $\approx$2. The fourth column summarizes the outlier probability of the different data sets. The values are computed from the average of the outlier probabilities of all data points in a given set, as predicted by \texttt{JAGS}. The data of \citet[][]{schmid97} have an average outlier probability of 72\%, in agreement with the elevated reduced $\chi^2$ found by \citet[][]{coc15} for this set. 

The same data sets were analyzed in both \citet[][]{coc15} and in the present work, and the same systematic uncertainties were adopted in both studies. Therefore, it is interesting to investigate the main reason for the significant differences, mentioned above, that are obtained in the analysis of the data of \citet[][]{casella02} and \citet[][]{bystritsky08}. We performed a series of tests and found that neither inclusion or omission of the lowest lying data point in \citet[][]{casella02}, nor the use of the correct or incorrect center-of-mass energies in \citet[][]{bystritsky08} (see Footnote~\ref{fn:data}) had an effect on our derived normalization factors listed in the top part of Table~\ref{tab:tabdpg}. These changes in the data sets are too small to affect the analysis. We also performed a test by using Gaussian instead of lognormal likelihoods for the data points and obtained again results in agreement with those listed in Table~\ref{tab:tabdpg}. This is not surprising because, with few exceptions, the data points have relatively small error bars, implying that a Gaussian closely approximates the lognormal likelihood (Section~\ref{sec:nongauss}). We thus conclude that the significant differences obtained presently and previously regarding the data sets of \citet[][]{casella02} and \citet[][]{bystritsky08} are caused by the adoption of a Bayesian model in our work as opposed to using a traditional method employed by \citet[][]{coc15}.

%We repeated the Bayesian analysis by using Gaussian instead of lognormal likelihoods for the data points and obtained consistent results within uncertainties. This is not surprising because in the analyzed data sets there is only a single data point (i.e., the datum at the lowest bombarding energy in \citet[][]{casella02}) with a mean-value-to-error ratio in excess of a factor of $3$. All the other data points have relatively small error bars, implying that a Gaussian is a close approximation to the lognormal likelihood (Section~\ref{sec:nongauss}).

The lower part of Table~\ref{tab:tabdpg} compares the present and previous values for the scale factor of the theoretical model results, and the astrophysical S-factor at zero energy. Our Bayesian analysis verifies the results reported by \citet[][]{coc15}, both for the recommended values and the magnitude of the uncertainties. Our zero-energy S-factor also agrees with the value presented in \citet[][]{adelberger11}, although our uncertainty ($3.7$\%) is smaller by a factor of $2$. The analysis by \citet[][]{adelberger11} was performed using a $\chi^2$ minimization and assuming a quadratic parameterization of the S-factor. The zero-energy S-factor presented in \citet[][]{xu13} has a much larger uncertainty ($19$\%) compared to all other recent values. It was obtained, using a potential model, from a standard $\chi^2$ fit in conjunction with a ``fit-by-eye'' technique.

In summary, completely independent methods of analysis provide comparable results. But unlike the $\chi^2$  minimization applied previously, the Bayesian technique provides consistent answers without the need to resort to Gaussian assumptions and other approximations. Thermonuclear reaction rates will be presented in Section~\ref{sec:rrdpg}.

\subsection{$^{3}$He($^{3}$He,2p)$^{4}$He}\label{sec:he3he3}
The $^{3}$He($^{3}$He,2p)$^{4}$He reaction represents the third and final step of the pp1 chain. The competition of this process with the $^{3}$He($\alpha$,$\gamma$)$^{7}$Be reaction determines the relative neutrino fluxes that originate from the pp and pep reactions (pp1 chain) compared to the $^{7}$Be and $^{8}$B decays (pp2 and pp3 chains). The S-factor ratio for $^{3}$He($^{3}$He,2p)$^{4}$He and $^{3}$He($\alpha$,$\gamma$)$^{7}$Be enters directly in the calculation of the solar neutrino energy losses, and thus impacts the relationship between the photon luminosity and the total energy production of the Sun \citep[][]{adelberger11}.

\begin{figure*}[!htb]
\centering{\includegraphics[width=2.0\columnwidth]{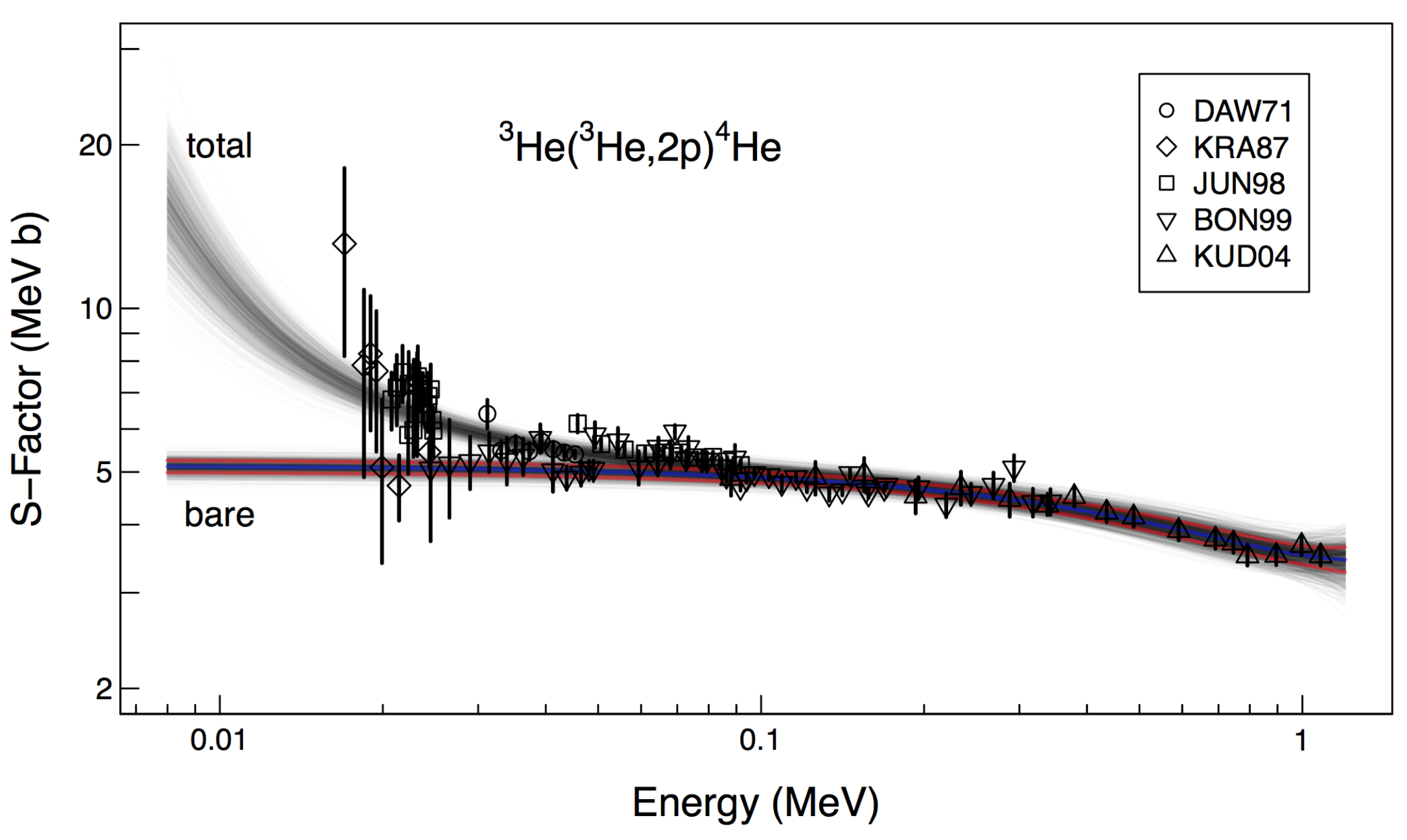}}
\caption{
Astrophysical S-factor versus center-of-mass energy for the $^{3}$He($^{3}$He,2p)$^{4}$He reaction. The symbols show the data of \citet[][]{dwarakanath71} (circles); \citet[][]{krauss87} (diamonds); \citet[][]{junker98} (squares); \citet[][]{bonetti99} (inverted triangles); \citet[][]{kudomi04} (triangles). The error bars refer to ($1\sigma$) statistical uncertainties only. The lines have the following meaning: (grey shaded area) credible S-factors, obtained from the output of the \texttt{JAGS} model, where each line corresponds to one specific set of model parameters; (blue) median (50th percentile) of all credible lines; (red) 16th and 84th percentiles of all credible lines. The upper grey lines are calculated using a quadratic expansion of the bare nucleus S-factor, multiplied by an exponential factor that takes into account laboratory electron screening (see text). The lower grey lines represent the bare nucleus S-factor (i.e., without electron screening corrections). Two data points (not shown) at the lowest measured bombarding energies of \citet[][]{bonetti99}, which have mean-value-to-standard-deviation ratios in excess of a factor of $3$, have been omitted from our analysis (see text).
}\label{fig:he3he3}
\end{figure*}

The astrophysical S-factor of $^{3}$He($^{3}$He,2p)$^{4}$He was recently evaluated by \citet[][]{adelberger11}. As discussed above and in that work, a reliable estimation of the astrophysical S-factor requires separate knowledge of statistical and systematic uncertainties. This information is reported in four studies. The quoted systematic uncertainties are 4.5\% \citep[][]{krauss87}, 3.7\% \citep[][]{junker98}, 5.7\% \citep[][]{bonetti99}, and 3.8\% \citep[][]{kudomi04}. To these data we added one more study \citep[][]{dwarakanath71} for which we could infer separate statistical (4\%-7\%) and systematic uncertainties (8.2\%) based on the information provided. Our reasoning is discussed in more detail in Appendix~\ref{app:he3he3}. The data adopted in the present work are displayed as black symbols in Figure~\ref{fig:he3he3}, where the displayed error bars refer to ($1\sigma$) {\it statistical} uncertainties only. 

We disregarded the two data points from \citet[][]{bonetti99} at center-of-mass energies of $16.50$~keV and $17.46$~keV, with reported S-factor values of $7.70\pm7.70$~MeVb and $5.26\pm5.26$~MeVb, respectively, for the reasons given in Section~\ref{sec:nongauss}. Only a single event was observed at each of these two energies, and the quoted errors refer to statistical uncertainties only (see their Table I). It is not difficult to include such data in a Bayesian model {\it if their probability densities were known}. Since this information is not provided by \citet[][]{bonetti99}, we cannot include these two data points with very large errors. They should also not be included in a traditional ($\chi^2$ minimization) analysis. \citet[][]{bonetti99} and \citet{adelberger11} do not mention if they included these two data points or not.

The S-factor fit to the data was performed in previous work \citep[][]{bonetti99,adelberger11} by using the expressions
\begin{eqnarray}
S(E) = S_{bare}(E)~e^{\pi \eta \left(\frac{U_e}{E}\right)} \\
S_{bare}(E) = S(0) + S^{\prime}(0)E + \frac{1}{2}S^{\prime\prime}(0)E^2
\end{eqnarray}
where $S_{bare}$ is the bare nucleus S-factor that is not influenced by electron screening, $S(0)$ is the S-factor at zero center-of-mass energy, $S^{\prime}(0)$ and $S^{\prime\prime}(0)$ are the first and second energy derivatives of the S-factor at zero energy, and $U_e$ is the electron-screening potential energy. This expression adequately describes the total measured S-factor, $S(E)$, at energies below $1.1$~MeV.

Our Bayesian model has nine parameters. Five of these are the normalizations of the individual data sets. The highly informative priors for these parameters are computed using the systematic uncertainty factors quoted above. The other parameters are $S(0)$, $S^{\prime}(0)$, $S^{\prime\prime}(0)$, and $U_e$. We assume non-informative priors for these parameters, i.e., normal probability densities located at zero with large values for the standard deviations. The distributions for $S(0)$ and $U_e$ were truncated at zero energy since both the S-factor  and the electron-screening potential energy are positive quantities. Other choices of priors gave consistent results. 

We generated random samples using three independent Markov chains, each of length 75,000 (without burn-in). Results are shown in Figure~\ref{fig:he3he3}. The two sets of credible lines display the S-factors with (upper grey lines) and without (lower grey lines) electron screening corrections. All other lines have the same meaning as in Figure~\ref{fig:dpg}.

%%%%%%%%%%%%%%%%%%%%%%%%%%%%%%%%%%%%%%%%%%%%%%%%%%%%%%%%%%%%%%%%%%%%%%%%%%%%%%%%%%
\onecolumngrid
\begin{deluxetable}{cccccccc}
\tablecolumns{8}
\tablewidth{2.0\columnwidth}
\tablecaption{Results for the $^3$He($^3$He,2p)$^4$He Reaction\label{tab:tabhe3he3}}
\tablehead{
\multicolumn{2}{c}{Data}  & \colhead{} & \multicolumn{2}{c}{Present\tablenotemark{a}} &   \colhead{}   & \multicolumn{2}{c}{} \\
\cline{1-2} \cline{4-5} \\
\colhead{Ref.\tablenotemark{b}} & \colhead{n\tablenotemark{c}}   & \colhead{} & \colhead{norm\tablenotemark{d}}    & \colhead{outlier\tablenotemark{e}} & \colhead{} & \colhead{}    & \colhead{}}
\startdata
Dwa 71 &  17  &  & 1.000$^{+0.032}_{-0.031}$ & 18\% & &     & \\ 
Kra 87  & 47   &  & 0.977$^{+0.022}_{-0.021}$ & 67\% & &      & \\
Jun 98   & 25   & & 1.040$^{+0.023}_{-0.022}$ & 14\% & &     & \\
Bon 99  & 8 &     & 0.955$^{+0.044}_{-0.040}$ & 58\% & &      & \\
Kud 04  &  8  &   & 0.991$^{+0.023}_{-0.022}$ & 31\% & &     & \\ 
\\
\hline
\\
\multicolumn{2}{c}{Quantity} & & \multicolumn{2}{c}{Present\tablenotemark{a,f}} & & Previous\tablenotemark{g} & Previous\tablenotemark{h} \\
\\
\hline
\\
\multicolumn{2}{l}{$S(0)$~(MeVb):}                                 & & \multicolumn{2}{c}{$5.14^{+0.14}_{-0.13}$}      & & $5.32 \pm 0.08$   &  $5.32 \pm 0.23$ \\
\multicolumn{2}{l}{$S^\prime(0)$~(b):}                             & & \multicolumn{2}{c}{$-2.69^{+0.54}_{-0.54}$}    & & $-3.7 \pm 0.6$    &  $-6.44 \pm 1.29$ \\
\multicolumn{2}{l}{$S^{\prime\prime}(0)$~(b/MeV):}         & & \multicolumn{2}{c}{$2.14^{+0.89}_{-0.91}$}     & & $3.9 \pm 1.0$   &  $30.7 \pm 12.2$ \\
\multicolumn{2}{l}{$U_e$~(eV):}                                      & & \multicolumn{2}{c}{$325^{+47}_{-48}$}            & & $294 \pm 47$    &  $280 \pm 70$ \\
\multicolumn{2}{l}{$S(E_0$ $=$ $21.94$~keV$)$~(MeVb)\tablenotemark{i}:}      & & \multicolumn{2}{c}{$5.08^{+0.14}_{-0.13}$}      & &       &  $5.11 \pm 0.22$ \\
\enddata
\tablenotetext{a}{Uncertainties are derived from the 16th, 50th, and 84th percentiles of the probability density (posterior).}
\tablenotetext{b}{Reference labels: Dwa 71 \citep[][]{dwarakanath71}; Kra 87 \citep[][]{krauss87}; Jun 98 \citep[][]{junker98}; Bon 99 \citep[][]{bonetti99}; Kud04 \citep[][]{kudomi04}.}
\tablenotetext{c}{Number of data points in a given set.}
\tablenotetext{d}{Normalization of each data set, taking into account the reported systematic uncertainties (see text); the values correspond to 16th, 50th, and 84th percentiles of the probability density (posterior).}
\tablenotetext{e}{Probability that the data set is an outlier; computed from the average outlier probability of a given set.}
\tablenotetext{f}{Fit is valid for center-of-mass energies of $\leq 1.1$~MeV; each quoted value is marginalized over all other parameters (see text).}
\tablenotetext{g}{From $\chi^2$ minimization of \citet[][]{bonetti99} (see their Table~II).}
\tablenotetext{h}{From $\chi^2$ minimization of \citet[][]{adelberger11}, using their quadratic representation of the bare nucleus S-factor (see their Table~II).}
\tablenotetext{i}{S-factor at an energy of $21.94$~keV, corresponding to a temperature of $15.5$~MK at the center of the Sun.}
\end{deluxetable}
\twocolumngrid

%%%%%%%%%%%%%%%%%%%%%%%%%%%%%%%%%%%%%%%%%%%%%%%%%%%%%%%%%%%%%%%%%%%%%%%%%%%%%%%%%%

Our results are listed in Table~\ref{tab:tabhe3he3} and they are compared to recently published values obtained using traditional statistics ($\chi^2$ minimization). The top part of the table displays the normalization factors (``norm'') of each data set, taking into account the reported systematic uncertainties (see above). The fourth column summarizes the outlier probability of the different data sets, which is computed from the average of the outlier probabilities of all data points in a given set, as predicted by \texttt{JAGS}. We find the largest outlier probabilities for the data of \citet[][]{krauss87} (67\%) and \citet[][]{bonetti99} (58\%).

The lower part of Table~\ref{tab:tabhe3he3} compares the present and previous values for the S-factor expansion coefficients, $S(0)$, $S^\prime(0)$, $S^{\prime\prime}(0)$, and the electron screening potential, $U_e$. Notice that each of the listed present values is marginalized over all other parameters (see Appendix~\ref{sec:regression}). Our results agree with those of \citet[][]{bonetti99} within uncertainties. However, our  values for $S^\prime(0)$ and $S^{\prime\prime}(0)$ disagree with those reported by \citet[][]{adelberger11}. Since their value of $S(0)$ agrees with our result, we conclude that the disagreement for the other parameters is caused by the significantly smaller bombarding energy range analyzed by \citet[][]{adelberger11}, i.e., $0$ $-$ $350$~keV, compared to $0$ $-$ $1.1$~MeV in the present work. Our uncertainty of $2.6$\% for the zero-energy S-factor is much smaller than the value of $9.4$\% reported by \citet[][]{xu13}, who obtained the S-factor using a phenomenological nuclear reaction model and a $\chi^2$ minimization.

Furthermore, \citet[][]{adelberger11} report an S-factor of $S_{Adelberger}(E_0)$ $=$ $5.11$ $\pm$ $0.22$~MeVb at the Gamow peak ($E_0$ $=$ $21.94$~keV) for the Sun's central temperature ($T$ $=$ $15.5$~MK), corresponding to an uncertainty of $4.3$\%. Our result is $S_{present}(E_0)$ $=$ $5.08^{+0.14}_{-0.13}$~MeVb, corresponding to a significantly smaller uncertainty of $2.7$\%. Thermonuclear reaction rates will be presented in Section~\ref{sec:rrhe3he3}.

\subsection{$^{3}$He($\alpha$,$\gamma$)$^{7}$Be}\label{sec:3heag}
The detection of solar neutrinos has entered a precision era, enabling the measurement of neutrino fluxes with a total uncertainty of about $3$\% $-$ $5$\% by various neutrino detectors \citep[][]{aharmin13,smy13,bellini14}. The measured neutrino fluxes can be used to probe the solar core and test solar models, provided that the relevant thermonuclear reaction rates are accurately known. Since the $^{3}$He($\alpha$,$\gamma$)$^{7}$Be reaction competes with $^{3}$He($^{3}$He,2p)$^{4}$He, it determines the number of $^7$Be and $^8$B neutrinos originating from the pp2 and pp3 chains. The $^{3}$He($\alpha$,$\gamma$)$^{7}$Be reaction also plays a prominent role in big bang nucleosynthesis. While the primordial abundances of d, $^3$He, and $^4$He predicted by standard big bang models are in reasonable agreement with those from observation, the models overproduce the primordial abundance of $^7$Li by a factor of $\approx$3. This ``$^7$Li problem'' is among the unsolved mysteries in astrophysics \citep[][]{iocco09}. Most of the $^7$Li in the early universe is produced as $^7$Be, by the $^{3}$He($\alpha$,$\gamma$)$^{7}$Be reaction, and decays subsequently via electron capture to $^7$Li. Although a new determination of the $^{3}$He($\alpha$,$\gamma$)$^{7}$Be rate does not appear to solve the $^7$Li problem, it is nevertheless desirable to determine a reliable rate of this reaction.

\begin{figure*}[!htb]
\centering{\includegraphics[width=2.0\columnwidth]{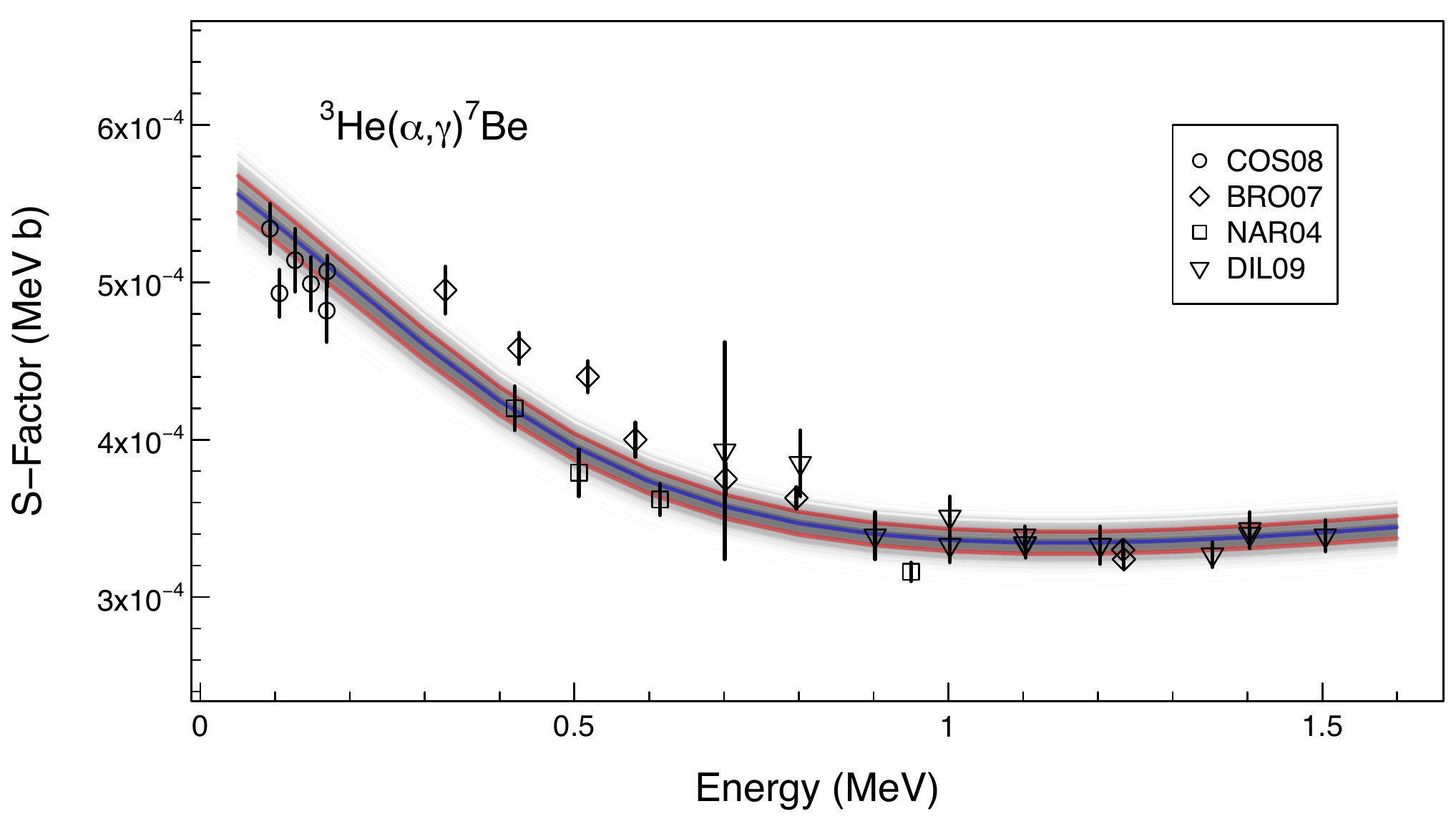}}
\caption{
Astrophysical S-factor versus center-of-mass energy for the $^{3}$He($\alpha$,$\gamma$)$^{7}$Be reaction. The symbols show the data of \citet[][]{costantini08} (circles); \citet[][]{brown07} (diamonds); \citet[][]{narasingh04} (squares); \citet[][]{dileva09} (inverted triangles). The error bars refer to ($1\sigma$) statistical uncertainties only. The lines have the following meaning: (grey shaded area) credible S-factors, obtained from the output of the \texttt{JAGS} model, where each line corresponds to one specific set of model parameters; (blue) median (50th percentile) of all credible lines; (red) 16th and 84th percentiles of all credible lines. The credible lines are calculated from the theoretical S-factor of \citet[][]{neff11}, multiplied by a scale factor that is a parameter in the Bayesian model. Notice the linear scale compared to the previous figures, for better comparison with other $^{3}$He($\alpha$,$\gamma$)$^{7}$Be S-factor plots published recently.
}\label{fig:he3ag}
\end{figure*}

Many groups have measured the $^{3}$He($\alpha$,$\gamma$)$^{7}$Be reaction using various experimental strategies. For summary discussions, see \citet[][]{adelberger11,bordeanu13,deboer14}. Similar to the procedure in \citet[][]{adelberger11}, we adopt a subset of all published measurements for the present analysis. First, we only consider those studies that provide separate statistical and systematic uncertainties. This excludes all measurements performed before the year 2000. Second, we focus on the center-of-mass energy region below $1.6$~MeV. Third, we only consider those experiments that directly measure the total cross section, i.e., via activation or recoil detection. We exclude prompt $\gamma$-ray data, since these studies rely so far on computed rather than measured corrections for $\gamma$-ray angular correlation effects. Based on these selection criteria, four data sets remain: \citet[][]{brown07,narasingh04,dileva09,costantini08}, which were labeled ``Seattle'', ``Weizmann'', ``ERNA'', ``LUNA'', respectively, in \citet[][]{adelberger11}. For systematic uncertainties, we adopt 3.0\% \citep[][]{brown07}, 5.1\% \citep[][]{narasingh04}, 5.0\% \citep[][]{dileva09}, and 3.1\% \citep[][]{costantini08}. Notice that \citet[][]{adelberger11} adopt a systematic uncertainty of only 2.2\% for the data of \citet[][]{narasingh04}. However, this value applies to their highest energy data point only, while the other three data points  have considerably higher systematic uncertainties (4.1\% $-$ 7.1\%). We adopt here the average value. The data adopted in the present work are displayed as black symbols in Figure~\ref{fig:he3ag}, where the displayed error bars refer to ($1\sigma$) {\it statistical} uncertainties only. 

Several different strategies have been employed in the past to fit the experimental S-factor data for the $^{3}$He($\alpha$,$\gamma$)$^{7}$Be reaction, including potential models \citep[][]{tombrello63}, parametrized analytical functions \citep[][]{cyburt08}, resonating-group methods \citep[][]{kajino86}, and R-matrix approaches \citep[][]{deboer14}. In this work, we will focus on three microscopic models. The first is the resonating-group study of \citet[][]{kajino86}, which has been used in several previous investigations. In this model, the nuclear system is characterized by antisymmetrized wave functions describing the relative motion of two clusters. The required phenomenological nucleon-nucleon interactions were tuned to reproduce the properties of bound and scattering states within the restricted cluster model space. The second is the calculation of \citet[][]{nollett01}, which employed accurate nucleon-nucleon potentials. The bound states were computed using the variational Monte Carlo method, while the relative motion of the nuclei in the initial state was described by one-body wave functions generated from the intercluster potential A of \citet[][]{kim81}. The third is the {\it ab initio} model of \citet[][]{neff11}, which employed realistic interactions to solve the many-body problem using a large model space. The latter work found that the assumption of a predominant external capture, which was commonly adopted in most previous studies, is not that well satisfied. Similar to the discussion in Section~\ref{sec:dpg}, we assume that these models adequately describe the shape of the S-factor, but that the absolute scale of each model S-factor is determined by a fit to the data. In the following, we first discuss and quote results obtained using the model of \citet[][]{neff11}. Subsequently, we use the models of \citet[][]{kajino86} and \citet[][]{nollett01} to estimate the model uncertainty for the extrapolation of the S-factor to low energies, where no data exist. We obtained the theoretical S-factors for all three models from the original authors as numerical tables, which were directly implemented into \texttt{JAGS}. Tests showed that this procedure caused negligible errors in the linear interpolation between grid points.

Our Bayesian model has five parameters. Four of these are the normalizations of the individual data sets. The highly informative priors for these parameters are computed using the systematic uncertainty factors quoted above. The fifth parameter is the common scaling factor by which the nuclear model results have to be multiplied to fit the data. We assume a non-informative prior for the latter parameter, i.e., a normal probability density with a location of zero and a standard deviation of $100$. The distribution was truncated at zero since the scaling factor must be a positive quantity. Other choices of priors gave very similar results. We generated random samples using three independent Markov chains, each of length 75,000 (without burn-in). Results are shown in Figure~\ref{fig:he3ag}, where the lines have the same meaning as in Figure~\ref{fig:dpg}. 

Our numerical results are listed in Table~\ref{tab:tabhe3ag} and are compared to recently published values obtained using different methods. The top part of the table displays the normalization factors (``norm'') of each data set, taking into account the reported systematic uncertainties (see above). The fourth column summarizes the outlier probability of the different data sets, which is computed from the average of the outlier probabilies of all data points in a given set, as predicted by \texttt{JAGS}. We obtain the largest average outlier probability for the data of \citet[][]{brown07} (81\%).

The lower part of Table~\ref{tab:tabhe3ag} compares the present and previous values for the S-factor at zero energy, $S(0)$. From fitting the data using the model of \citet[][]{neff11}, we find $S(0)_{present}$ $=$ $(5.72 \pm 0.12)$ $\times$ $10^{-4}$~MeVb, representing an uncertainty of $2.1$\%. Our result agrees within the quoted uncertainties with those of \citet[][]{adelberger11}, who used in their analysis the same data sets as we did. Notice, however, that \citet[][]{adelberger11} employed an analytic function that approximated the S-factor of \citet[][]{nollett01} ``to better than $0.3$\%, on average''. In contrast, we directly used the original numerical tables of \citet[][]{neff11} and there was no need for an approximation. Compared to \citet[][]{adelberger11}, our uncertainty in $S(0)$ from fitting the data is smaller by a factor of $\approx$ $2$. It is also interesting that our value for $S(0)$ disagrees with the R-matrix result of \citet[][]{deboer14}, $S(0)_{deBoer}$ $=$ $(5.42 \pm 0.11)$ $\times$ $10^{-4}$~MeVb, where their quoted error is based on the data fit only. Their quoted mean value of $S(0)$ is lower by $5.5$\% compared to our result.

% at 15.5 MK Gamow peak is at E0=22.8954 keV

%%%%%%%%%%%%%%%%%%%%%%%%%%%%%%%%%%%%%%%%%%%%%%%%%%%%%%%%%%%%%%%%%%%%%%%%%%%%%%%%%%

\onecolumngrid
\begin{deluxetable}{cccccccc}
\tablecolumns{8}
\tablewidth{2.0\columnwidth}
\tablecaption{Results for the $^3$He($\alpha$,$\gamma$)$^7$Be Reaction\label{tab:tabhe3ag}}
\tablehead{
\multicolumn{2}{c}{Data}  & \colhead{} & \multicolumn{2}{c}{Present\tablenotemark{a}} &   \colhead{}   & \multicolumn{2}{c}{} \\
\cline{1-2} \cline{4-5}  \\
\colhead{Ref.\tablenotemark{b}} & \colhead{n\tablenotemark{c}}   & \colhead{} & \colhead{norm\tablenotemark{d}}    & \colhead{outlier\tablenotemark{e}} & \colhead{} & \colhead{}    & \colhead{}}
\startdata
Nar 04  & 4   & & 0.966$^{+0.026}_{-0.025}$ & 36\%  & &  &     \\
Bro 07 & 8    & & 1.031$^{+0.024}_{-0.023}$ & 81\%  & &  &   \\
Cos 08 & 6   & & 0.977$^{+0.022}_{-0.021}$ & 28\% & &  &     \\
DiL 09 &  15 & & 1.003$^{+0.022}_{-0.021}$ & 9.0\%  & &  &  \\ 
\\
\hline
\\
\multicolumn{2}{c}{Quantity} & & \multicolumn{2}{c}{Present\tablenotemark{a}} & & \multicolumn{2}{c}{Previous} \\
\\
\hline
\\
\multicolumn{2}{c}{scale factor\tablenotemark{f}:} & & \multicolumn{2}{c}{$0.964^{+0.021}_{-0.020}$} & & \multicolumn{2}{c}{} \\
\multicolumn{2}{c}{S(0) (MeVb):} & & \multicolumn{2}{c}{$(5.72\pm0.12)\times10^{-4}$\tablenotemark{g}} & & \multicolumn{2}{c}{$(5.6\pm0.2)\times10^{-4}$\tablenotemark{h}} \\
\multicolumn{2}{c}{} & & \multicolumn{2}{c}{} & & \multicolumn{2}{c}{$(5.42\pm0.11)\times10^{-4}$\tablenotemark{i}} \\
\enddata
\tablenotetext{a}{Uncertainties are derived from the 16th, 50th, and 84th percentiles of the probability density (posterior).}
\tablenotetext{b}{Reference labels: Nar 04 \citep[][]{narasingh04}; Bro 07 \citep[][]{brown07}; Cos 08 \citep[][]{costantini08}; DiL 09 \citep[][]{dileva09}.}
\tablenotetext{c}{Number of data points in a given set.}
\tablenotetext{d}{Normalization of each data set, taking into account the reported systematic uncertainties (see text).}
\tablenotetext{e}{Probability that the data set is an outlier; computed from the average of the outlier probabilities of all data points in a given set.}
\tablenotetext{f}{Best estimate for the scale factor of the theoretical S-factor from \citet[][]{neff11}.}
\tablenotetext{g}{Uncertainty from data fit using the theoretical S-factor from \citet[][]{neff11} only; an additional ``theory uncertainty'' of $0.12\times10^{-4}$~MeVb is found when different theoretical models are used (see text).}
\tablenotetext{h}{From \citet[][]{adelberger11}; the original S-factor quoted in that work is $S(0)$ $=$ $0.56$ $\pm$ $0.02$(exp) $\pm$ $0.02$(theo)~keVb, where the latter uncertainty contribution was obtained for a range of theoretical models.}
\tablenotetext{i}{From \citet[][]{deboer14}; the original S-factor quoted in that work is $S(0)$ $=$ $0.542$ $\pm$ $0.011$(MC fit) $\pm$ $0.006$(model) $^{+0.019}_{-0.011}$(phase shifts)~keVb; the first uncertainty was obtained from the data fit, the second from varying the background pole energies and the R-matrix channel radius, and the third from using different scattering data sets to define the phase shifts.}
\end{deluxetable}
\twocolumngrid
%%%%%%%%%%%%%%%%%%%%%%%%%%%%%%%%%%%%%%%%%%%%%%%%%%%%%%%%%%%%%%%%%%%%%%%%%%%%%%%%%%

For low energies, especially those pertaining to the solar Gamow peak, data do not exist and the S-factor must be extrapolated to compute the reaction rates. Therefore, past work has included a ``theory error'' that is based on the spread in $S(0)$ values obtained when different theoretical models are used to fit the data. In our case, we repeated our analysis using the theoretical model S-factors of \citet[][]{kajino86} and \citet[][]{nollett01}. With the theoretical model of \citet[][]{nollett01}, we find almost identical S-factors (mean value and uncertainties) compared to the model of \citet[][]{neff11}. The model of \citet[][]{kajino86} resulted in a similar uncertainty, but a mean value smaller by a factor of $2.2$\%. Using the full spread of $2.2$\% as an estimate for the ``theory error'', our result is $S(0)_{present}$ $=$ $(5.72$ $\pm$ $0.12(\mathrm{exp})$ $\pm$ $0.13(\mathrm{theo}))$ $\times$ $10^{-4}$~MeVb. This can be compared to $S(0)_{Adelberger}$ $=$ $(5.6$ $\pm$ $0.2(\mathrm{exp})$ $\pm$ $0.2(\mathrm{\mathrm{theo}}))$ $\times$ $10^{-4}$~MeVb and $S(0)_{deBoer}$ $=$ $(5.42$ $\pm$ $0.11(\mathrm{MC fit})$ $\pm$ $0.06(\mathrm{model})$ $^{+0.19}_{-0.11}(\mathrm{phase~shifts}))$ $\times$ $10^{-4}$~MeVb. Concerning the result of \citet[][]{deboer14}, the first uncertainty was obtained from the data fit, the second from varying the background pole energies and the R-matrix channel radius, and the third from using different scattering data sets to define the phase shifts. Thermonuclear reaction rates will be presented in Section~\ref{sec:rrhe3he3}.

\section{Bayesian reaction rates}\label{sec:rates}
The thermonuclear reaction rate per particle pair, $N_A \langle \sigma v \rangle$, can be written as \citep[][]{iliadis15}
\begin{equation}
\begin{split}
N_A \langle \sigma v \rangle & = \left(\frac{8}{\pi m_{01}}\right)^{1/2} \frac{N_A}{(kT)^{3/2}} \\
& \int_0^\infty e^{-2\pi\eta}\,S(E)\,e^{-E/kT}\,dE 
\label{eq:rate}
\end{split}
\end{equation}
where $m_{01}$ is the reduced mass of projectile and target, and $N_A$ is  Avogadro's constant; the product of Boltzmann constant, $k$, and plasma temperature, $T$, is numerically given by
\begin{equation}
kT = 0.086173324 \,T_9 \,\,\, \mbox{(MeV)}
\end{equation}
with the temperature, $T_9$, given in units of GK. 

For each set of parameters sampled by the MCMC algorithm, we calculate the reaction rates by numerical integration of Eq.~(\ref{eq:rate}) on a grid of $60$ temperatures between $1$~MK and $10$~GK. The resulting set of reaction rate values constitute the reaction rate probability density at a given temperature. Using this probability density, we follow the procedure recommended in \citet[][]{longland10} to compute a recommended rate (50th percentile), a high and low rate (16th and 84th percentiles, respectively), and the lognormal parameters, $\mu$ and $\sigma$, of the lognormal approximation of the total reaction rate. The rate factor uncertainty, $f.u.$, corresponding to a coverage probability of 68\%, is obtained from $f.u.$ $=$ $e^\sigma$ (see Equation~\ref{eq:lognormal}). Notice we directly compute the lognormal parameters from the expectation value and variance of all rate samples, $\ln(N_A \langle \sigma v \rangle_i)$, at a given temperature. This  ensures that the results can be directly incorporated into the STARLIB library \citep[][]{sallaska13}.

\begin{figure}[!htb]
\centering{\includegraphics[width=1.0\columnwidth]{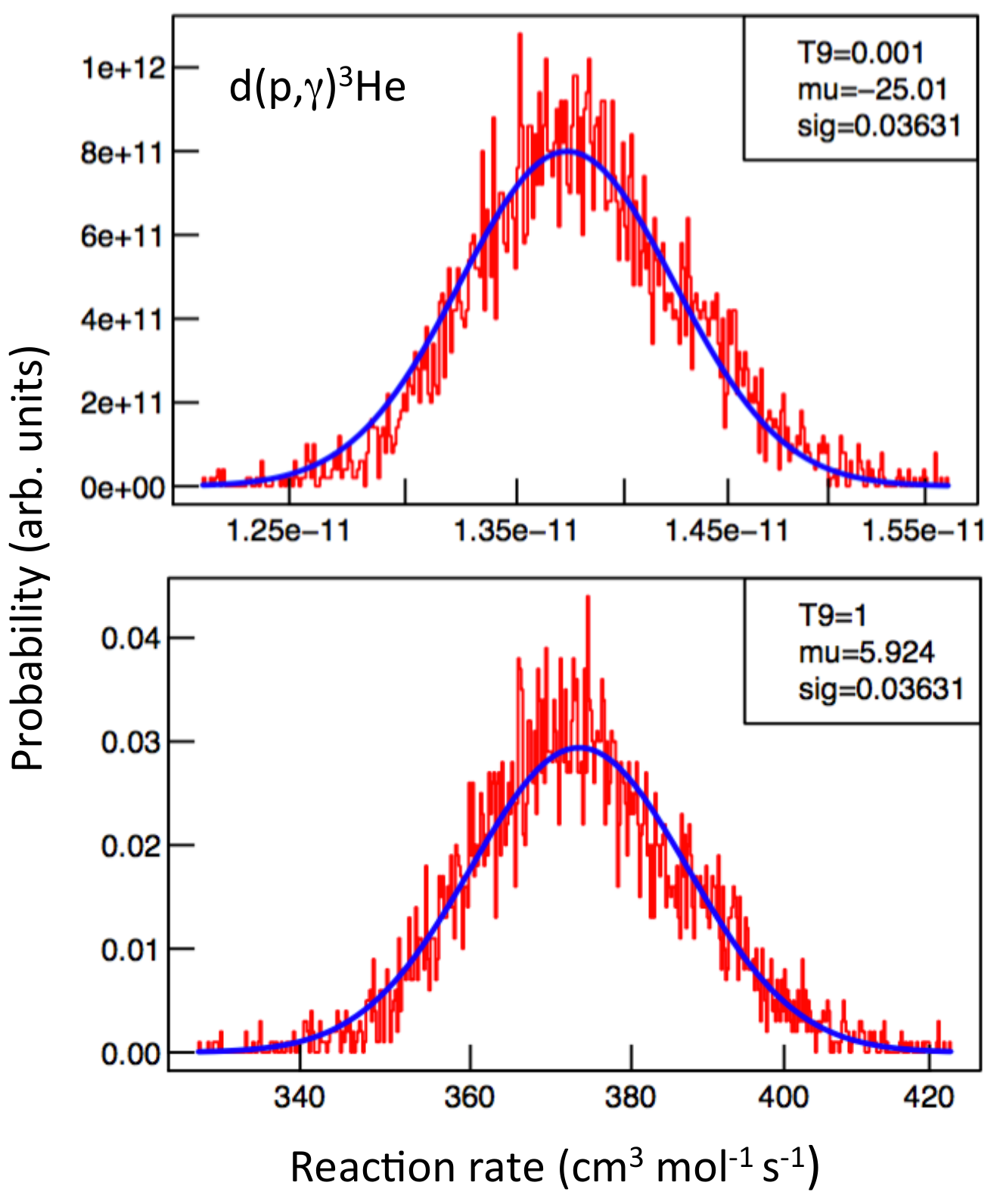}}
\caption{
Reaction rate probability densities for the d(p,$\gamma$)$^3$He reaction for two temperatures (``T9'' is in units of GK): (top) $T$=1~MK, near the range important for deuterium burning; (bottom) $T$= 1~GK, relevant for big bang nucleosynthesis. Rate samples (red histograms) are computed using the S-factor samples obtained from the Bayesian model. Blue curves represent  lognormal approximations, where the lognormal parameters $\mu$ (``mu'') and $\sigma$ (``sig'') are directly calculated from the expectation value and variance of all rate samples, $\ln(N_A \langle \sigma v \rangle_i)$, at a given temperature.
}\label{fig:ratedpg}
\end{figure}

\subsection{Reaction rates for d(p,$\gamma$)$^{3}$He}\label{sec:rrdpg}
The present rates for the d(p,$\gamma$)$^3$He reaction, together with the corresponding factor uncertainties, are listed in columns 2 and 3 of Table~\ref{tab:rates}. The rate factor uncertainty is constant, $f.u.$ $=$ $3.7$\% (except at the highest temperatures; see below), since it is determined by a single parameter (i.e., the common scaling factor; Section~\ref{sec:dpg}). Reaction rate probability densities are shown in Figure~\ref{fig:ratedpg} for two selected temperatures, $T$ $=$ $1$~MK (top), near the range important for deuterium burning, and $T$ $=$ $1$~GK (bottom), relevant for big bang nucleosynthesis. The reaction rate samples (red histograms) are computed using the S-factor samples obtained from the Bayesian model (Section~\ref{sec:dpg}). The sampled rates are well represented by lognormal probability densities, shown as blue curves.

For temperatures of $T$ $\geq$ $8$~MK, our rates agree with the recently evaluated results of \citet[][]{coc15} within $1$\%. However, at lower temperatures, important for deuterium burning, our rates deviate strongly from the previous results. For example, at the lowest temperature, $T$ $=$ $1$~MK, our rates are larger by a factor of $\approx 300$ compared to those of \citet[][]{coc15}\footnote{This large difference has no impact on big bang nucleosynthesis; see below.}. The disagreement is explained by an erroneously assumed lower integration limit of $2$~keV in the previous work, which is too high for computing the reaction rates at the lowest temperatures. Our estimated reaction rate factor uncertainties are close to the values given previously at all temperatures. We integrate numerically the reaction rates only up to $2$~MeV, i.e., the highest center-of-mass energy for which we have theoretical S-factors from \citet[][]{marcucci05}. Since we may miss rate contributions at the highest temperatures, $T$ $\geq$ $5$~GK, we adopt in this region the values from \citet[][]{coc15}, which are shown in italics in Table~\ref{tab:rates}.

% \onecolumngrid

\onecolumngrid
\begin{deluxetable}{ccccccccc}
\tablewidth{0pt}
\tablecaption{Present recommended reaction rates.\tablenotemark{a}\label{tab:rates}}
\tablehead{
 & \multicolumn{2}{c}{d(p,$\gamma$)$^3$He} & & \multicolumn{2}{c}{$^3$He($^3$He,2p)$^4$He} & & \multicolumn{2}{c}{$^3$He($\alpha$,$\gamma$)$^7$Be}\\
\cline{2-3} \cline{5-6} \cline{8-9}
\\
\colhead{T (GK)} & \colhead{Rate\tablenotemark{b}} & \colhead{$f.u.$\tablenotemark{b}} &  & \colhead{Rate\tablenotemark{c}} & \colhead{$f.u.$\tablenotemark{c}} &  & \colhead{Rate\tablenotemark{d}} & \colhead{$f.u.$\tablenotemark{d}} 
}
\startdata
 0.001 	&  1.379E-11  &  1.037  & & 2.700E-41        & 1.025  & &  1.178E-47   & 1.024  \\
 0.002	&  1.906E-08  &  1.037  & & 1.694E-30        & 1.025  & &  2.300E-36   & 1.024  \\
 0.003	&  6.175E-07  &  1.037  & & 2.890E-25        & 1.025  & &  6.811E-31   & 1.024  \\
 0.004	&  5.464E-06  &  1.037  & & 5.721E-22        & 1.025  & &  1.911E-27   & 1.024  \\
 0.005	&  2.557E-05  &  1.037  & & 1.259E-19        & 1.025  & &  5.391E-25   & 1.024  \\
 0.006	&  8.262E-05  &  1.037  & & 7.683E-18        & 1.025  & &  3.975E-23   & 1.024  \\
 0.007	&  2.101E-04  &  1.037  & & 2.043E-16        & 1.025  & &  1.230E-21   & 1.024  \\
 0.008	&  4.529E-04  &  1.037  & & 3.052E-15        & 1.025  & &  2.083E-20   & 1.024  \\
 0.009	&  8.655E-04  &  1.037  & & 2.997E-14        & 1.025  & &  2.273E-19   & 1.024  \\
 0.010	&  1.510E-03  &  1.037  & & 2.140E-13        & 1.025  & &  1.778E-18   & 1.024  \\
 0.011	&  2.456E-03  &  1.037  & & 1.193E-12        & 1.025  & &  1.073E-17   & 1.024  \\
 0.012	&  3.773E-03  &  1.037  & & 5.451E-12        & 1.025  & &  5.266E-17   & 1.024  \\
 0.013	&  5.538E-03  &  1.037  & & 2.120E-11        & 1.025  & &  2.182E-16   & 1.024  \\
 0.014	&  7.825E-03  &  1.037  & & 7.209E-11        & 1.025  & &  7.859E-16   & 1.024  \\
 0.015	&  1.071E-02  &  1.037  & & 2.191E-10        & 1.025  & &  2.517E-15   & 1.024  \\
 0.016	&  1.427E-02  &  1.037  & & 6.051E-10        & 1.025  & &  7.291E-15   & 1.024  \\
 0.018	&  2.368E-02  &  1.037  & & 3.647E-09        & 1.025  & &  4.782E-14   & 1.024  \\
 0.020	&  3.662E-02  &  1.037  & & 1.711E-08        & 1.025  & &  2.413E-13   & 1.024  \\
 0.025	&  8.760E-02  &  1.037  & & 3.765E-07        & 1.024  & &  6.141E-12   & 1.024  \\
 0.030	&  1.702E-01  &  1.037  & & 3.957E-06        & 1.024  & &  7.209E-11   & 1.024  \\
 0.040	&  4.480E-01  &  1.037  & & 1.207E-04        & 1.024  & &  2.582E-09   & 1.024  \\
 0.050	&  8.922E-01  &  1.037  & & 1.359E-03        & 1.024  & &  3.258E-08   & 1.024  \\
 0.060	&  1.511E+00  &  1.037  & & 8.560E-03       & 1.024   & & 2.238E-07   &  1.024  \\
 0.070	&  2.304E+00  &  1.037  & & 3.705E-02       & 1.023   & & 1.038E-06   & 1.024  \\
 0.080	&  3.267E+00  &  1.037  & & 1.236E-01       & 1.023   & & 3.666E-06   & 1.024  \\
 0.090	&  4.395E+00  &  1.037  & & 3.414E-01       & 1.023   & & 1.062E-05   & 1.024  \\
 0.100	&  5.680E+00  &  1.037  & & 8.172E-01       & 1.023   & & 2.648E-05   & 1.024  \\
 0.110	&  7.115E+00  &  1.037  & & 1.749E+00      & 1.023   & &  5.876E-05  &  1.024  \\
 0.120	&  8.693E+00  &  1.037  & & 3.426E+00      & 1.023   & & 1.187E-04   & 1.024  \\
 0.130	&  1.041E+01  &  1.037  & & 6.240E+00      & 1.023   & & 2.224E-04   & 1.024  \\
 0.140 	&  1.225E+01  &  1.037  & & 1.070E+01      & 1.023   & & 3.913E-04   & 1.024  \\
 0.150	&  1.421E+01  &  1.037  & & 1.746E+01      & 1.022   & & 6.530E-04   & 1.024  \\
 0.160	&  1.629E+01  &  1.037  & & 2.729E+01      & 1.022   & & 1.042E-03   & 1.024  \\
 0.180	&  2.078E+01  &  1.037  & & 6.000E+01      & 1.022   & & 2.376E-03   & 1.024  \\
 0.200	&  2.567E+01  &  1.037  & & 1.180E+02      & 1.022   & & 4.818E-03   & 1.024  \\
 0.250	&  3.945E+01  &  1.037  & & 4.534E+02      & 1.022   & & 1.968E-02   & 1.024  \\
 0.300	&  5.510E+01  &  1.037  & & 1.255E+03      & 1.022   & & 5.698E-02   & 1.024  \\
 0.350	&  7.231E+01  &  1.037  & & 2.813E+03      & 1.022   & & 1.322E-01   & 1.024  \\
 0.400	&  9.084E+01  &  1.037  & & 5.448E+03      & 1.022   & & 2.631E-01   & 1.024  \\
 0.450	&  1.105E+02  &  1.037  & & 9.492E+03      & 1.023   & & 4.687E-01   & 1.024  \\
 0.500	&  1.311E+02  &  1.037  & & 1.526E+04      & 1.023   & &  7.677E-01  &  1.024  \\
 0.600	&  1.749E+02  &  1.037  & & 3.316E+04      & 1.024   & & 1.717E+00  &  1.024  \\
 0.700	&  2.215E+02  &  1.037  & & 6.120E+04      & 1.024   & & 3.239E+00  &  1.024  \\
 0.800	&  2.703E+02  &  1.037  & & 1.009E+05      & 1.025   & & 5.435E+00  &  1.024  \\
 0.900	&  3.211E+02  &  1.037  & & 1.534E+05      & 1.025   & &  8.380E+00 &  1.024  \\
 1.000	&  3.734E+02  &  1.037  & & 2.193E+05      & 1.026   & & 1.213E+01  &  1.024  \\
 1.250	&  5.101E+02  &  1.037  & & 4.430E+05      & 1.027   & & 2.519E+01  &  1.024  \\
 1.500	&  6.534E+02  &  1.037  & & {\it 7.96E+05}  & {\it 1.076}  & &  4.360E+01   &  1.024  \\
 1.750	&  8.017E+02  &  1.037  & & {\it 1.21E+06}  & {\it 1.077}  & &  6.717E+01   &  1.024  \\
 2.000	&  9.540E+02  &  1.037  & & {\it 1.70E+06}  & {\it 1.079}  & &   9.539E+01  &  1.024  \\
 2.500	&  1.268E+03  &  1.037  & & {\it 2.90E+06}  & {\it 1.081}  & &  {\it 1.705E+02}   & {\it 1.035} \\
 3.000	&  1.590E+03  &  1.037  & & {\it 4.32E+06}  & {\it 1.082}  & &  {\it 2.585E+02}   & {\it 1.035} \\
 3.500	&  1.916E+03  &  1.037  & & {\it 5.95E+06}  & {\it 1.081}  & &  {\it 3.602E+02}   & {\it 1.035} \\
 4.000	&  2.241E+03  &  1.037  & & {\it 7.75E+06}  & {\it 1.081}  & &  {\it 4.742E+02}   & {\it 1.035} \\
 5.000	&  {\it 2.905E+03}  &  {\it 1.040}  & & {\it 1.18E+07}  & {\it 1.079} & & {\it 7.351E+02}  & {\it 1.035} \\
 6.000	&  {\it 3.557E+03}  &  {\it 1.042}  & & {\it 1.63E+07}  & {\it 1.077} & & {\it 1.035E+03}  & {\it 1.035} \\
 7.000	&  {\it 4.194E+03}  &  {\it 1.044}  & & {\it 2.12E+07}  & {\it 1.073} & & {\it 1.370E+03}  & {\it 1.035} \\
 8.000	&  {\it 4.812E+03}  &  {\it 1.046}  & & {\it 2.63E+07}  & {\it 1.069} & & {\it 1.738E+03}  & {\it 1.035} \\
 9.000	&  {\it 5.410E+03}  &  {\it 1.047}  & & {\it 3.15E+07}  & {\it 1.067} & & {\it 2.135E+03}  & {\it 1.035} \\
10.000	&  {\it 5.988E+03}  &  {\it 1.049}  & & {\it 3.68E+07}  & {\it 1.063} & & {\it 2.558E+03}  & {\it 1.035} \\
\enddata
\tablenotetext{a}{In units of cm$^3$~mol$^{-1}$~s$^{-1}$. The values correspond to the {\it median} rate, i.e., the 50th percentile of the rate probability density. The rate factor uncertainty, $f.u.$, corresponding to a coverage probability of 68\%, is calculated from $f.u.$ $=$ $e^\sigma$, where $\sigma$ denotes the spread parameter of the lognormal approximation to the rate probability density (see Equation~\ref{eq:lognormal}).} 
\tablenotetext{b}{Values for $T$ $\geq$ $5$~GK, shown in italics, are adopted from \citet[][]{coc15} (see text).}
\tablenotetext{c}{Values for $T$ $\geq$ $1.5$~GK, shown in italics, are adopted from \citet[][]{angulo99} (see text).}
\tablenotetext{d}{Values for $T$ $\geq$ $2.5$~GK, shown in italics, are adopted from \citet[][]{kontos13} (see text).}
\end{deluxetable}
 
\twocolumngrid

It is straightforward to calculate the effect of the new reaction rate on 
the predicted primordial D/H ratio. Our mean value for the scale factor 
(Table~\ref{tab:tabdpg}) represents a $1$\% increase compared to \citet[][]{coc15}. This difference translates into a decrease of the central D/H value by only $0.32$\% \citep[][]{iocco09}, while the total uncertainty remains unchanged at $2$\%.

\subsection{Reaction rates for $^3$He($^3$He,2p)$^{4}$He}\label{sec:rrhe3he3}
\begin{figure}[!htb]
\centering{\includegraphics[width=1.0\columnwidth]{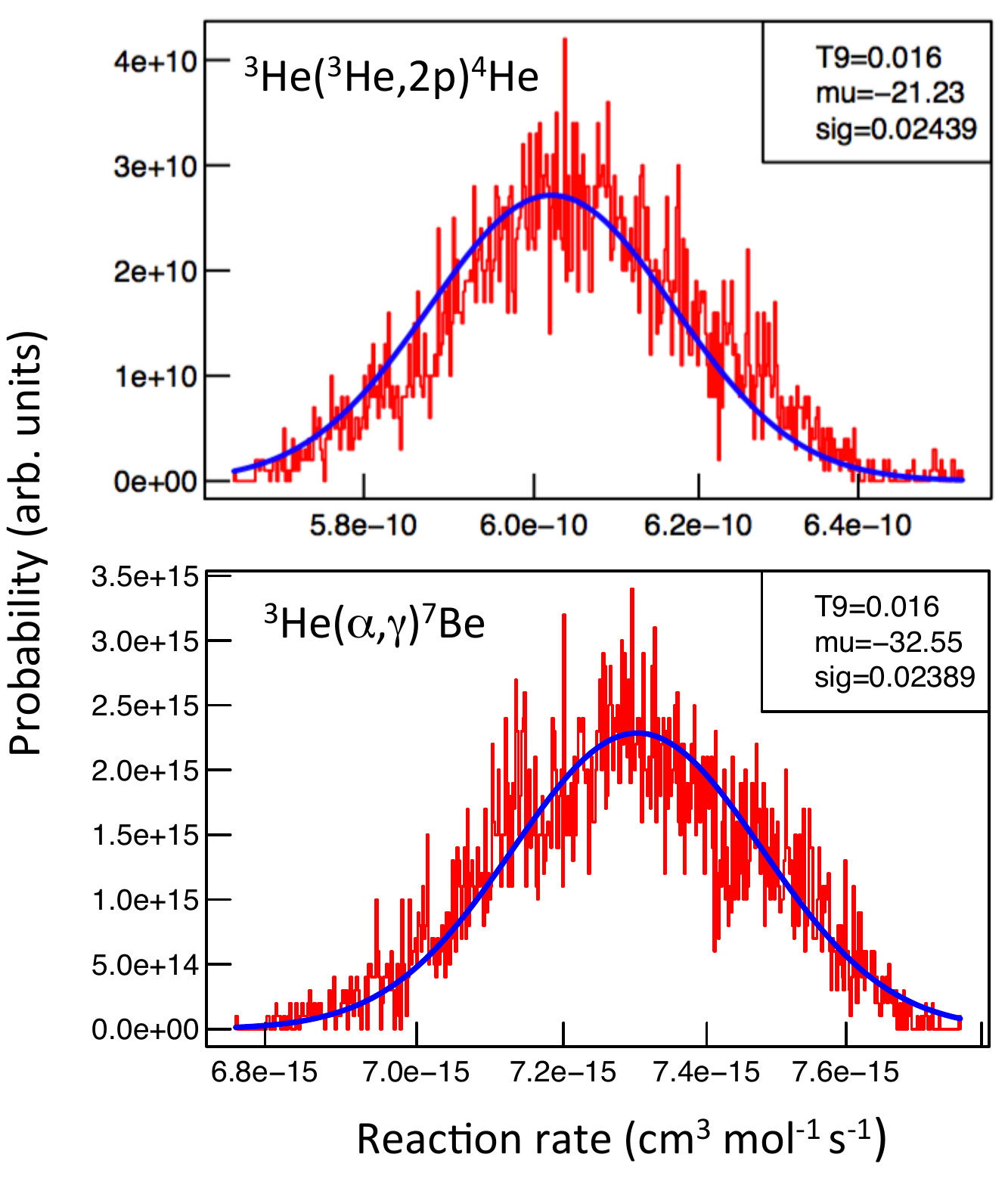}}
\caption{
Reaction rate probability density of (top) $^3$He($^3$He,2p)$^4$He, and (bottom) $^3$He($\alpha$,$\gamma$)$^7$Be, for the temperature at the Sun's center ($15.5$~MK). The rate samples (red histograms) are computed using the S-factor samples obtained from the Bayesian model. The blue curve represents a lognormal approximation, where the lognormal parameters $\mu$ (``mu'') and $\sigma$ (``sig'') are directly calculated from the expectation value and variance of all rate samples, $\ln(N_A \langle \sigma v \rangle_i)$, at a given temperature (``T9'' is in units of GK).
}\label{fig:ratehe3he3}
\end{figure}

The present rates for the $^3$He($^3$He,2p)$^{4}$He reaction, together with the corresponding factor uncertainties, are listed in columns 4 and 5 of Table~\ref{tab:rates}. The rate factor uncertainties amount to $2.2$\% $-$ $2.7$\% for temperatures of $T$ $\leq$ $1.25$~GK. The reaction rate probability density is shown in Figure~\ref{fig:ratehe3he3} (top) for the temperature at the Sun's center ($15.5$~MK). The reaction rate samples (red histograms) are computed using the S-factor samples obtained from the Bayesian model (Section~\ref{sec:he3he3}). The sampled rates are well represented by a lognormal probability density, shown as blue curve.

Although for temperatures of $T$ $\leq$ $1.25$~GK our rates agree with those of \citet[][]{angulo99}, our estimated reaction rate factor uncertainties are significantly smaller, by a factor of $\approx2.7$ (i.e., $\approx2.4$\% versus $\approx6.5$\%). Since we numerically integrate the reaction rates only up to $1.1$~MeV, we may miss rate contributions at the highest temperatures, $T$ $>$ $1.25$~GK. Therefore, we adopt in this region the values from \citet[][]{angulo99}, which are shown in italics in Table~\ref{tab:rates}. 

At temperatures relevant for the center of the Sun ($T$ $\approx$ $15.5$~MK), the $^7$Be and $^8$B solar neutrino fluxes approximately scale with the $S(0)$ value according to the relations $\phi_{\nu}^{^{7}Be}$ $\sim$ $S(0)^{-0.43}$ and $\phi_{\nu}^{^{8}B}$ $\sim$ $S(0)^{-0.40}$ \citep[Table~XV in][]{bahcall88}. Therefore, compared to the rate of \citet[][]{adelberger11}, our results translate into an increase in the $^7$Be and $^8$B solar neutrino fluxes by 1.5\% and 1.4\%, respectively. A more detailed analysis, incorporating our much reduced uncertainty in $S(0)$ (by factor of $2$), is beyond the scope of the present work. 

\subsection{Reaction rates for $^3$He($\alpha$,$\gamma$)$^{7}$Be}\label{sec:rr3heag}
The present rates for the $^3$He($\alpha$,$\gamma$)$^{7}$Be reaction, together with the corresponding factor uncertainties, are listed in columns 6 and 7 of Table~\ref{tab:rates}. The rate factor uncertainty amounts to $2.4$\% for temperatures of $T$ $\leq$ $2.0$~GK. The reaction rate probability density is shown in Figure~\ref{fig:ratehe3he3} (bottom) for the temperature at the Sun's center ($15.5$~MK). The reaction rate samples (red histograms) are computed using the S-factor samples obtained from the Bayesian model (Section~\ref{sec:3heag}). The sampled rates are well represented by a lognormal probability density, shown as blue curve.

For the Sun's central temperature, the present reaction rates barely agree with the R-matrix results of \citet[][]{deboer14} within the quoted uncertainties (corresponding to 68\% probability density intervals). However, our recommended rate is larger by 6.0\%. At big bang temperatures ($\approx$ $1$~GK), our result is in good agreement with the rate of \citet[][]{deboer14}. As already mentioned, we fit the S-factor data up to an energy of $1.6$~MeV. Therefore, we can only compute the reaction rates up to a temperature of $2.0$~GK. For higher temperatures, we adopt the values from \citet[][]{kontos13}, which are shown in italics in Table~\ref{tab:rates}.

At temperatures relevant for the center of the Sun ($T$ $\approx$ $15.5$~MK), the $^7$Be and $^8$B solar neutrino fluxes approximately scale with the $S(0)$ value according to the relations $\phi_{\nu}^{^{7}Be}$ $\sim$ $S(0)^{0.86}$ and $\phi_{\nu}^{^{8}B}$ $\sim$ $S(0)^{0.81}$ \citep{bahcall88}. Therefore, compared to the rate of \citet[][]{deboer14}, our results translate into an increase in the $^7$Be and $^8$B solar neutrino fluxes by 4.7\% and 4.5\%, respectively.

The $^3$He($\alpha$,$\gamma$)$^7$Be rate is the major nuclear physics source of uncertainty for the prediction of the primordial $^7$Li abundance. The $^7$Li/H ratio varies almost linearly with the reaction rate \citep[][]{iocco09}. The study of primordial nucleosynthesis by \citet[][]{coc15} adopted the rate of \citet[][]{deboer14}, which agrees with our result at big bang temperatures  within a few percent. Therefore, we expect only minor modifications to the predicted primordial $^7$Li/H ratio. Such small variations are negligible compared to the factor of $3$ discrepancy between predicted and observed primordial $^7$Li/H ratios, and thus are not relevant for the $^7$Li problem.

\section{Summary}\label{sec:summary}
We discussed astrophysical S-factors and reaction rates based on Bayesian statistics, and developed a framework that incorporates robust parameter estimation, systematic effects, and non-Gaussian uncertainties. Unlike the $\chi^2$ minimization applied previously, the Bayesian technique provides consistent answers without the need to resort to Gaussian assumptions and other frequently applied approximations. The method is used to estimate the d(p,$\gamma$)$^3$He, $^3$He($^3$He,2p)$^4$He, and $^3$He($\alpha$,$\gamma$)$^7$Be S-factors and reaction rates, important for deuterium burning, solar neutrinos, and big bang nucleosynthesis.

For the d(p,$\gamma$)$^3$He reaction, our analysis verifies the results reported by Coc et al. (2015), both for the recommended values and the magnitude of the uncertainties. Our zero-energy S-factor also agrees with the value presented in Adelberger et al. (2011), although our uncertainty is smaller by a factor of $\approx$ $2$. The zero-energy S-factor presented in Xu et al. (2013) has a much larger uncertainty (19\%) compared to all other recently published values. Our reaction rate factor uncertainty is 3.7\% for all temperatures below $5$~GK. Compared to \citet[][]{coc15}, our reaction rate at big bang temperatures is larger by about $1$\%. This translates into a decrease in the primordial D/H value by only $0.32$\%, while the total uncertainty remains unchanged at $2$\%.

For the $^3$He($^3$He,2p)$^4$He reaction, our results agree with those of Bonetti et al. (1999) within uncertainties. However, our parameter values for $S^{\prime}(0)$ and $S^{\prime\prime}(0)$ disagree with those reported by Adelberger et al. (2011). Our uncertainty of 2.6\% for the zero-energy S-factor is much smaller than the value of 9.4\% reported by Xu et al. (2013). Furthermore, Adelberger et al. (2011) report an S-factor of $S_{Adelberger}(E_0)$ $=$ $5.11\pm0.22$~MeVb at the Gamow peak ($E_0$ $=$ $21.94$~keV) for the Sun's central temperature ($T$ $=$ $15.5$~MK), corresponding to an uncertainty of 4.3\%. Our result is $S_{present}(E_0)$ $=$ $5.08\pm0.14$~MeVb, corresponding to a smaller uncertainty of 2.7\%. Compared to the reaction rate of \citet[][]{adelberger11}, our results translate into an increase in the $^7$Be and $^8$B solar neutrino fluxes by 1.5\% and 1.4\%, respectively.

For the $^3$He($\alpha$,$\gamma$)$^7$Be reaction, we find $S(0)_{present}$ $=$ $(5.72 \pm 0.12)$ $\times$ $10^{-4}$~MeVb, representing an uncertainty of 2.1\%, from fitting the data using the {\it ab initio} model of Neff (2011). Our result agrees within the quoted uncertainties with that of Adelberger et al. (2011). However, compared to the latter work, our uncertainty in $S(0)$ from fitting the data is smaller by a factor of $\approx$ $2$. Also, our value for $S(0)$ disagrees with the R-matrix result of deBoer et al. (2014). Their quoted mean value of $S(0)$ is lower by 5.5\% compared to our result. This translates into an increase in the $^7$Be and $^8$B solar neutrino fluxes by 4.7\% and 4.5\%, respectively.

% nuclear model inherent uncertainties?? \\
% suggestions: model comparison!! what fit function to use! \\

\acknowledgments
We would like to thank Stefano Andreon, Jack Dermigny, Ryan Fitzgerald, Jordi Jos\'e, Richard Longland, Thomas Neff, and Ken Nollett for their input and feedback. This work was supported in part by NASA under the Astrophysics Theory Program grant 14-ATP14-0007 and the Theoretical and Computational Astrophysics Networks grant NNX14AB53G, U.S. DOE under Contract No. DE-FG02-97ER41041, and NSF under the Software Infrastructure for Sustained Innovation grant 1339600 and grant PHY 08-022648 for the Physics Frontier Center ``Joint Institute for Nuclear Astrophysics -- Center for the Evolution of the Elements'' (JINA-CEE). F.X.T. acknowledges sabbatical support from the Simons Foundation.

\software{JAGS \citep{plummer03}, R \citep{rcore15}}

%%%%%%%%%%%%%%%%%%%%%%%%%%%%%%%%%%%%%%%%%%%%%
 
\appendix
\section{Bayesian inference}\label{sec:inference}
Bayesian methods have revolutionized many scientific fields, including archeology, ecology, genetics, linguistics, political science, and psychology. A brief historical account and a comparison of traditional  (``frequentist") and Bayesian statistics can be found in \citet[][]{brooks03}. An introduction to Bayesian inference in physics \citep[][]{vontoussaint11} and a textbook on Bayesian methods for the physical sciences \citep[][]{andreon15} have been published recently.

Denoting with $p(A|B)$ the probability that ``proposition A is true given that proposition B is true", we can write the product rule of elementary logic as
\begin{mathletters}
\begin{eqnarray}
p(A \wedge B) &=& p(A|B)p(B) \\
p(B \wedge A) &=& p(B|A)p(A) \label{eqn:bayes1}
\end{eqnarray}
\end{mathletters}
Since $A \wedge B$ $=$ $B \wedge A$, solving for $p(A|B)$ yields the general form of Bayes' theorem
\begin{equation}
p(A|B) = \frac{p(B|A)p(A)}{p(B)} \label{eqn:bayes2}
\end{equation}
The above expression applies to any kind of proposition. When applied to experimental data and continuous model parameters, it can be written as \citep[][]{kruschke15}
\begin{equation}
p(\theta|D) = \frac{p(D|\theta)p(\theta)}{p(D)} = \frac{p(D|\theta)p(\theta)}{\int{d\theta p(D|\theta)p(\theta)}} \label{eqn:bayes3}
\end{equation}
The factor $p(D|\theta)$ is the {\it likelihood} function, the same as in traditional (``frequentist") statistics, and denotes the {\it probability that the data, $D$, were obtained assuming given values for the model parameters, $\theta$}. Since in most cases more than one parameter is involved in a given model, $\theta$ denotes the complete set of model parameters, ($\theta_1$, $\theta_2$,..., $\theta_n$). The factor $p(\theta)$ is called the {\it prior}, which represents our state of knowledge before seeing the data. The product of likelihood and prior defines the factor $p(\theta|D)$, called the {\it posterior}. The denominator, called the {\it evidence}, is a normalization factor representing the product of likelihood and prior, integrated over all values of the parameters, $\theta$. All of the factors entering in Bayes' theorem represent probability densities. 

Equation~\ref{eqn:bayes3} shows that the traditional maximum likelihood estimate, obtained by maximizing the likelihood function, generally differs from the posterior estimate because of the presence of the prior, $p(\theta)$. The maximum likelihood estimate is often mistakenly interpreted as ``the most probable estimate given the data". This is incorrect since in frequentist statistics the model parameters are not random variables. Their true values are unknown. In Bayesian statistics, on the other hand, the model parameters are random variables and the posterior provides directly the information we seek, that is, {\it the probability of a given set of model parameters given the data}. Bayesian inference is used for parameter estimation, value prediction, and model selection. We will be concerned in this work with parameter estimation and value prediction only. In that case, only the numerator on the right-hand side of Equation~\ref{eqn:bayes3} is of interest.  

In a Bayesian analysis, it is important to compare posterior inferences under different reasonable choices of prior distributions. The posterior will be insensitive to the choice of prior when the sample size is large. However, when the sample size is small, the prior distribution becomes more important. Prior distributions range from ``non-informative'', e.g., a uniform density between two reasonable limits, to ``highly informative", e.g., when fairly precise information is available for a given parameter \citep[][]{gelman02}. In this work, we will explore uniform, Gaussian, and lognormal distributions as prior densities. 

Without the use of numerical algorithms, the Bayesian method discussed so far is only applicable to very simple problems, involving few parameters, for which analytical solutions exist. The main reason for the wide adoption of Bayesian techniques in many scientific fields is that the random sampling of the posterior can be performed numerically over many parameter dimensions using Markov chain Monte Carlo (MCMC) algorithms \citep[][]{metropolis53,hastings70,geyer11}. 

A Markov chain is a random walk, where a transition from state $i$ to state $j$ is independent (``memory-less'') of how state $i$ was populated. The fundamental theorem of Markov chains states that for a very long random walk the proportion of time (i.e., probability) the chain spends in some state $j$ is independent of the initial state it started from. This set of limiting, long random walk, probabilities is called the stationary (or equilibrium) distribution of the Markov chain. Consequently, when a Markov chain is constructed with a stationary distribution equal to the posterior, $p(\theta|D)$, the samples drawn at every step during a sufficiently long random walk will closely approximate the posterior density. Several related algorithms (e.g., Metropolis, Metropolis-Hastings, Gibbs) are known to solve this problem numerically. The combination of Bayes' theorem and MCMC algorithms allows for computing models that are too difficult to estimate using traditional statistical methods.

In this work we employ the program \texttt{JAGS} (``Just Another Gibbs Sampler'') for the analysis of Bayesian models using Markov chain Monte Carlo sampling \citep[][]{plummer03}. Specifically, we will employ the \texttt{rjags} package that works directly with \texttt{JAGS} within the R language \citep[][]{rcore15}. Running a \texttt{JAGS} model refers to generating random samples from the posterior distribution of model parameters. This involves the definition of the model, likelihood, and priors, as well as the initialization, adaptation, and monitoring of the Markov chain. 

Two major issues that need to be tested in a Bayesian analysis are the mixing and convergence of the Markov chains, and the sensitivity of the results to the priors. The former is achieved using suitable diagnostic tools \citep[][]{gelman92,geweke92,raftery92}, while the latter can be investigated by comparing posterior inferences under different reasonable choices of prior distribution. 

\section{A simple example: linear regression}\label{sec:regression}
To illustrate these ideas with a simple example, we apply the Bayesian method to the problem of linear regression. A comprehensive treatment of regression using Bayesian statistics can be found in \citet[][]{gelman07}. 

Suppose we aim to fit a straight line to some data, $D$, given by $\left\{ x_1, x_2,...,x_n \right\}$, $\left\{ y_1, y_2,...,y_n \right\}$, and $\left\{ \epsilon_1, \epsilon_2,...,\epsilon_n \right\}$, with $x_i$, $y_i$, and $\epsilon_i$ the independent variable, dependent variable, and the error in the dependent variable, respectively. For simplicity, we will assume no error on the independent variable. The linear relationship between variables satisfies 
\begin{equation}
y^\prime = \alpha + \beta x
\end{equation}
but $y^\prime$ cannot be observed directly. Instead, we observe the quantity
\begin{equation}
y_i = y_i ^\prime+ \epsilon_i
\label{eq:noise}
\end{equation}
If we further assume for this simple example that the errors, $\epsilon_i$, are Gaussian random variables with standard deviations of $\sigma_i$, the likelihood function for all data points is
\begin{equation}
p(D | \alpha \beta) = \prod\limits_{i=1}^n \frac{1}{\sigma_i \sqrt{2\pi}} e^{
{-\frac{(y_i - [\alpha + \beta x_i])^2}{2\sigma_i^2}}
} 
\end{equation}
which represents a product of normal distributions, each with a mean of $y_i^\prime$ and a standard deviation of $\sigma_i$. In abbreviated form, we may write symbolically for each data point 
\begin{equation}
y_i \sim \mathcal{N} (y_i^\prime,\sigma_i^2)
\end{equation}
implying that its value is sampled from a normal distribution with a mean equal to the true value, $y_i^\prime$, and a variance of $\sigma_i^2$.

\begin{figure}[!htb]
\centering{\includegraphics[width=1.0\columnwidth]{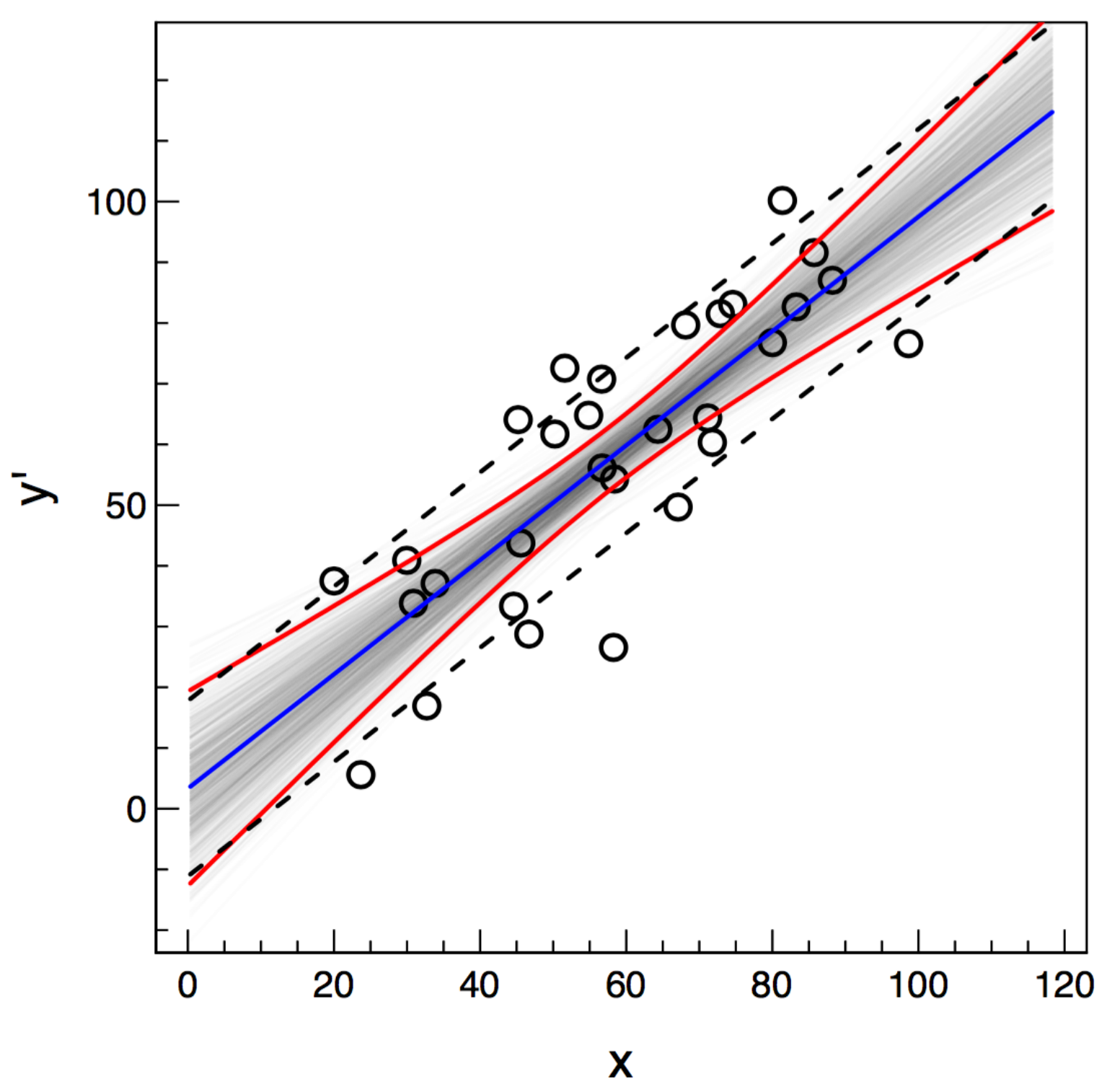}}
\caption{
Artificially generated data set, assuming a linear relationship between independent and dependent variables, with an intercept of $\alpha$ $=$ $0.0$, a slope of $\beta$ $=$ $1.0$, and a scatter of $\sigma$ $=$ $15$. (Grey shaded area) Region of credible regression lines, obtained from the output of the \texttt{JAGS} model (see Figure~\ref{fig:jagsout}); each line corresponds to one specific set of model parameters ($\alpha$, $\beta$, $\sigma$). (Blue line) Median (50th percentile) of all credible regression lines. (Red lines) 16th and 84th percentiles of all credible regression lines. (Dashed lines) Median regression line plus or minus the mean value of the sampled scatter parameter $\sigma$.
}\label{fig:linreg}
\end{figure}

A specific example is displayed in Figure~\ref{fig:linreg}. The artificial data shown as open circles have been generated under the following assumptions: (i) first, $30$ $x$-values are sampled from a uniform probability density in the range from $x_i$ $=$ $20$ to $x_f$ $=$ $120$; (ii) the corresponding $y$-values are computed using the linear relationship $y^\prime$ $=$ $0.0 + 1.0 x$, i.e., an intercept of $0.0$ and a slope of $1.0$; (iii) a noise contribution to each $y^\prime$-value is sampled from a normal probability density with a mean of $0$ and a standard deviation of $15$; (iv) the observed value, $y$, is obtained by adding the noise contribution to the true (but unobserved) value of $y^\prime$, according to Equation~\ref{eq:noise}. In other words, in this simple example we assume the scatter is represented by the same distribution for all $30$ data points.

We now analyze this hypothetical data set using Bayesian inference and are particularly interested to see if the analysis recovers the input parameters used to generate the data, i.e., the slope, the intercept, and the magnitude of the noise. The \texttt{JAGS} model, in symbolic notation\footnote{Unlike R, Fortran, or C, \texttt{JAGS} is a {\it declarative} language, i.e., the syntax provided here is a model declaration, and does not define a set of computational steps to be run sequentially. At compilation, the model declaration syntax is turned into a set of instructions that would correspond to a program in the conventional sense, but this is never seen by the user. Therefore, the precise order in which statements are given in the model declaration is unimportant.}, is set up as follows:
\begin{verbatim}
# LIKELIHOOD
obsy[i] ~ dnorm(y[i], pow(sigmay, -2))
y[i] = alpha + beta * obsx[i]
# PRIORS
alpha ~ dnorm(0.0, pow(100, -2))
beta ~ dnorm(0.0, pow(100, -2))
sigmay ~ dunif(0, 200)
\end{verbatim}

\begin{figure*}[!htb]
\centering{\includegraphics[width=2.0\columnwidth]{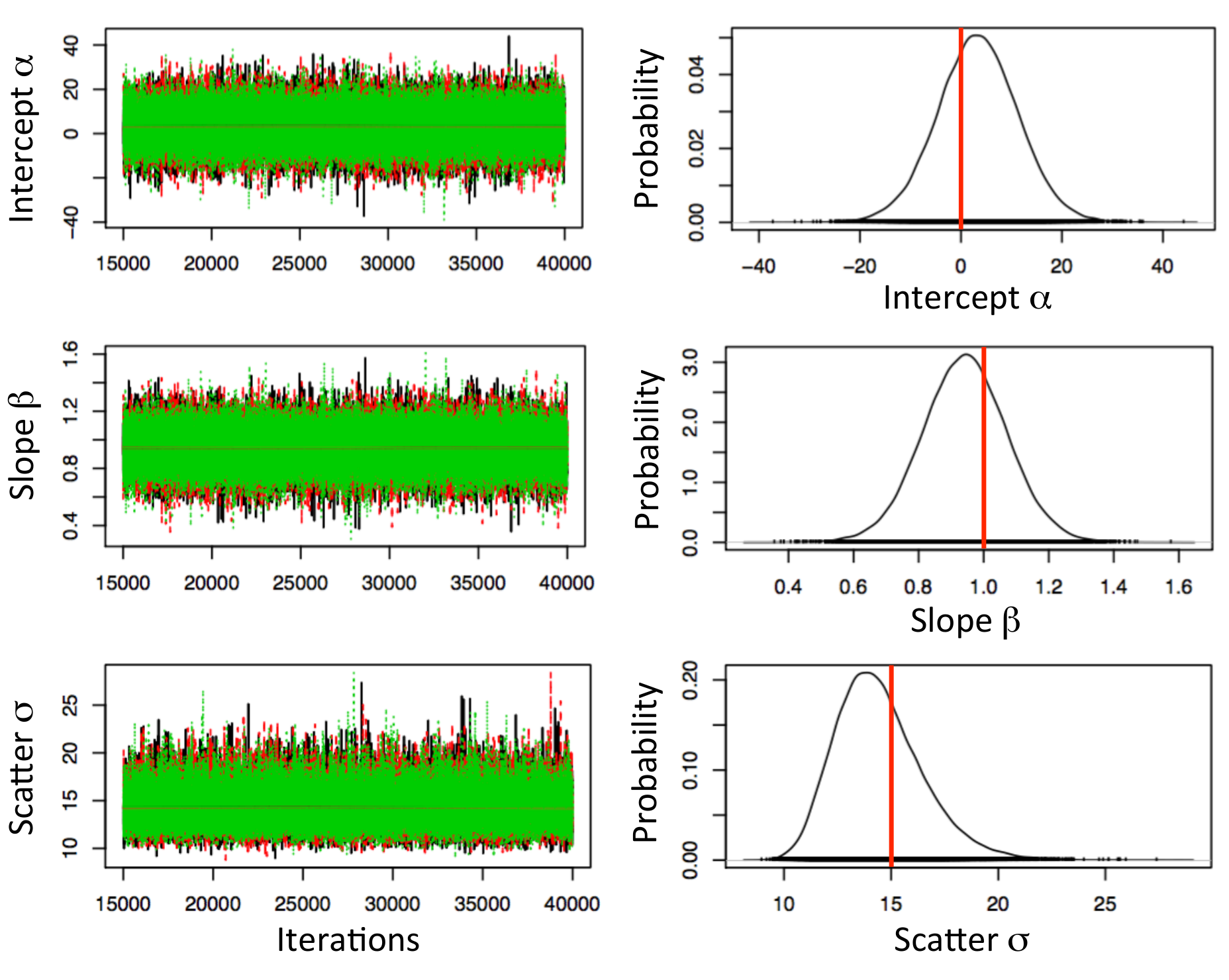}}
\caption{
Output of \texttt{JAGS} model for the data displayed in Figure~\ref{fig:linreg}. (Left side) Sampled values of parameters $\alpha$, $\beta$, and $\sigma$ versus sample number for three independent Markov chains, shown in different color, each of length $25,000$. The first $15,000$ samples were discarded to ensure equilibrium (``burn-in''). (Right side) Corresponding posterior densities. The vertical red lines indicate the parameter values used to generate the artificial data set.
}\label{fig:jagsout}
\end{figure*}

The third line states our model, i.e., a linear relationship between the unobserved (true) $y$-value and the $x$-value. The second line describes the likelihood for each data point, $i$. The ``$\sim$'' symbol stands for ``distributed as'' or ``sampled from''. Thus we sample for each data point the observed value, $y^\prime$ (\texttt{obsy}), from a normal probability density with a mean given by the true value, $y$, and a standard deviation of $\sigma$ (\texttt{sigmay}). The \texttt{pow()} command appears because \texttt{JAGS} requires as input the precision, $\tau$ $\equiv$ $1/\sigma^2$, instead of the standard deviation, $\sigma$. The three parameters of our model are the intercept ($\alpha$), the slope ($\beta$), and the scatter ($\sigma$). Next, each of these parameters require a prior probability density. For the slope and intercept we adopt normal distributions with a mean of zero and a standard distribution of $100$, i.e., very broad and slowly declining priors. For the standard deviation of the noise, we assume a uniform prior between values of $0$ and $200$.

The output of the \texttt{JAGS} model is displayed in Figure~\ref{fig:jagsout}. The panels on the left-hand side show $3 \times 25,000$ samples of $\alpha$, $\beta$, and $\sigma$, for three independent Markov chains (indicated by different colors in each panel). The first $15,000$ samples were discarded to ensure that the chains have achieved equilibrium (``burn-in''). It is apparent that the scatter is uniform and the chains are well mixed. The panels on the right-hand side display the corresponding posterior densities for $\alpha$, $\beta$, and $\sigma$. The 16th, 50th (median), and 84th percentiles extracted from the posteriors are $\alpha$ $=$ $3.3$ $\pm$ $7.9$, $\beta$ $=$ $0.94$ $\pm$ $0.13$, and $\sigma$ $=$ $14.2_{-1.8}^{+2.2}$, and thus the orginal values used to generate the data set, shown as vertical red lines, are  recovered within uncertainty. Notice that the panels shown on the right-hand side represent ``marginalized'' posterior densities, i.e., each distribution was obtained by integrating out the other two parameters. It would be inappropriate to use the full widths of the posteriors for estimating regression lines. Such a procedure would overestimate the uncertainties because of parameter correlations. 

Credible regression lines, calculated using the sampled values for the intercept ($\alpha$) and slope ($\beta$), marginalized over the scatter ($\sigma$), form the grey shaded area in Figure~\ref{fig:linreg}. A few interesting observations can be made. First, the density of grey lines decreases with increasing distance from the data points. Second, the width of the credible region is smallest in the middle of the data set, near $x$ $\approx$ $70$, and increases towards lower and higher $x$-values. Both observations agree with expectation, since the uncertainties should increase in regions devoid of data. We can quantify the credible region by computing suitable percentiles of $y$-values on a grid of $x$-values. The 50th percentile is shown as a blue line, whereas the 16th and 84th percentiles are displayed as red lines. Therefore, at any given $x$-value, there is a 68\% probability that the true (but usually unknown) $y$-value is located between the two red lines. 

The two dashed lines in Figure~\ref{fig:linreg} correspond to the 50th percentile, plus or minus the mean of the sampled values of the scatter, $\sigma$. The region between the two dashed lines indicates the most likely location of new data points acquired under the same conditions as the data shown. 

We will now provide the \texttt{JAGS} implementation of systematic uncertainties, robust regression, and non-Gaussian statistical uncertainties, discussed in Section~\ref{sec:regression2}, for the example of linear regression. The inclusion of a systematic normalization uncertainty of a factor of $1.10$  can be accomplished by:
\begin{verbatim}
# LIKELIHOOD
obsy[i] ~ dnorm(z[i], pow(erobsy[i], -2))
z[i] = n.factor * y[i]
y[i] = alpha + beta * obsx[i]
# PRIORS
alpha ~ dnorm(0.0, pow(100, -2))
beta ~ dnorm(0.0, pow(100, -2))
## lognormal density
n.factor ~ dlnorm(logmu, pow(logsigma, -2))
logmu = log(1.0)
logsigma = log(1.1)
\end{verbatim}
The first line contains the experimental statistical uncertainty, \texttt{erobsy}, for each individual data point, $i$, instead of assuming the same error as we did above. The second line includes the normalization factor (\texttt{n.factor}) into the likelihood function. The last three lines contain the information about the prior for the systematic uncertainty: \texttt{n.factor} is sampled from a lognormal density (\texttt{dlnorm}) with parameters of logmu=0 and logsigma=log(1.1), where \texttt{log} denotes the natural logarithm.

Robust regression can be implemented into our simple example as:

\begin{verbatim}
# LIKELIHOOD
obsy[i] ~ dnorm(y[i], pow(corr.er[i], -2))
p.alt[i] ~ dcat(p[])
corr.er[i] = erobsy[i] * phi[p.alt[i]]
y[i] = alpha + beta * obsx[i]
# PRIORS
alpha ~ dnorm(0.0, pow(100, -2))
beta ~ dnorm(0.0, pow(100, -2))
# if measured errors are correct:
phi[1] = 1
# if measured errors are overoptimistic:
phi[2] ~ dunif(1, 50)
p[1] ~ dunif(0, 1)
p[2] = 1 - p[1]
\end{verbatim}
First, a two-element vector \texttt{phi[]} is defined: for phi[1]=1 it is assumed that the reported uncertainty for a given datum, $i$, is correct; phi[2] is sampled from a uniform distribution between $1$ and some higher value (e.g., $50$ in this example). For example, if p[1]=0.2 and p[2]=0.8, then \texttt{dcat()} will return values of $1$ or $2$ with a probability of $20\%$ and $80\%$, respectively, each time \texttt{dcat()} is called. If p.alt[]=1, we obtain phi[1]=1, and the observed uncertainty for a given datum is assumed to be correct. If, on the other hand, p.alt[]=2, then the reported uncertainty of a given datum is multiplied by a factor of phi[2]. The MCMC sampling quantifies the outlier probability for a given datum by counting the number of times the indices 1 (no outlier) or 2 (outlier) have been called.

To account for the lognormal likelihood of the data analyzed here, we can replace the first line in the last \texttt{JAGS} code shown above,
\begin{verbatim}
# LIKELIHOOD
obsy[i] ~ dnorm(y[i], pow(corr.er[i], -2))
\end{verbatim}
which is appropriate for a Gaussian likelihood function, by 
\begin{verbatim}
# LIKELIHOOD
obsy[i] ~ dlnorm(ylg[i], 
        pow(corrlg.er[i],-2))                 
ylg[i] = log(y[i])-0.5*log(1+
       (pow(corr.er[i],2)/pow(y[i],2))) 
corrlg.er[i] = sqrt(log(1+
       (pow(corr.er[i],2)/pow(y[i],2))))   
\end{verbatim}
The quantities \texttt{ylg[i]} and \texttt{corrlg.er[i]} denote the lognormal parameters, $\mu$ and $\sigma$, respectively, of datum $i$ (Equations~\ref{eqn:lognorm1} and \ref{eqn:lognorm2}).

\section{Nuclear data}
\subsection{The d(p,$\gamma$)$^3$He reaction}\label{app:dpg}
The data for the d(p,$\gamma$)$^{3}$He reaction analyzed in the present work have recently been evaluated by \citet[][]{coc15}. We adopt the data listed in their Appendix B, with two exceptions. First, the energies in \citet[][]{bystritsky08} have been misinterpreted by \citet[][]{coc15}. The correct center-of-mass energies, used in the present work, of the three data points are $8.28$~keV, $9.49$~keV, and $10.10$~keV (see Footnote~\ref{fn:data}). Second, the data point at the lowest measured bombarding energy of \citet[][]{casella02} has a mean-value-to-standard-deviation ratio in excess of a factor of $3$ and has been omitted in our analysis for the reasons given in Section~\ref{sec:nongauss}.

\subsection{The $^3$He($^3$He,2p)$^4$He reaction}\label{app:he3he3}
\subsubsection{The data of \citet{kudomi04} and \citet{bonetti99}}
The S-factors from \citet{kudomi04} are taken from their Table II and are reproduced in Table~\ref{tab:kud}, which only lists statistical uncertainties. The authors state that the sum of the systematic uncertainties for the S-factor is 3.8\%.  They also re-evaluate the systematic uncertainties reported in the experiments of \citet{krauss87} and \citet{junker98}, and obtain 5.5\% and 3.7\%, respectively. 

The S-factors listed in Table I of \citet{bonetti99}  are reproduced in Table~\ref{tab:bon}.

\begin{deluxetable}{cc|cc}
\tablecaption{Data of \citet{kudomi04}.\label{tab:kud}} 
\tablewidth{\columnwidth}
\tablehead{
$E_{c.m.}$   & $S \pm \Delta S_{\mathrm{stat}}\tablenotemark{a}$  & $E_{c.m.}$   & $S \pm \Delta S_{\mathrm{stat}}\tablenotemark{a}$  \\
(MeV)  &  (MeVb) &  (MeV)  &  (MeVb) 
} 
\startdata
0.0312  & 6.40 $\pm$ 0.39 &0.0393  & 5.69 $\pm$ 0.25 \\
0.0331  & 5.48 $\pm$ 0.22 &0.0413  & 5.51 $\pm$ 0.18 \\
0.0352  & 5.62 $\pm$ 0.21 &0.0433  & 5.43 $\pm$ 0.14 \\
0.0373  & 5.46 $\pm$ 0.20 &0.0453  & 5.39 $\pm$ 0.09 \\
\enddata
\tablenotetext{a}{Systematic uncertainty: 3.8\%.}
\end{deluxetable}
\begin{deluxetable}{cccc}
\tablecaption{Data of \citet{bonetti99}.\label{tab:bon}} 
\tablewidth{\columnwidth}
\tablehead{
$E_{c.m.}$   & $S$  &  ${\Delta}S_\mathrm{stat}$  &  ${\Delta}S_\mathrm{sys}$  \\
 (MeV)  &  (MeVb) &   (MeVb) &   (MeVb) 
}
\startdata
0.01650  & 7.70 &  7.70 & 0.49 \\  
0.01699  & 13.15 & 4.98 & 0.83 \\
0.01746  & 5.26  & 5.26 & 0.33 \\
0.01846  & 7.86  & 2.97  & 0.47 \\
0.01898   & 8.25  & 2.29 & 0.48 \\
0.01946   & 7.67  & 2.22  & 0.44 \\
0.01993   & 5.10  & 1.70 & 0.29 \\
0.02143   & 4.72 &  0.65  & 0.26 \\
0.02337   & 7.31  & 0.63 & 0.39 \\
0.02436   & 5.44 &  0.34 & 0.28 \\
\enddata
\end{deluxetable}

\subsubsection{Data of \citet{junker98} and \citet{krauss87}}
The S-factors obtained from Table I of \citet{junker98} are presented in Table~\ref{tab:jun} with statistical uncertainties only, and supersede the preliminary results reported by \citet{Arp96}. \citet{junker98} note that the systematic uncertainty (one standard deviation) includes uncertainties in the gas target pressure (1\%), beam power (3\%), detection efficiency (2\%), beam energy resolution, and beam energy loss (10\%). \cite{kudomi04} re-evaluated their total systematic uncertainty and find a value of 3.7\%.

The S-factors of \citet{krauss87} are extracted from their Table I and are listed in Table~\ref{tab:kra} with statistical uncertainties only. Systematic uncertainties of 3.0\% and 3.4\% from the excitation function normalization and the absolute cross section scale, respectively, have to be considered as well.

\begin{deluxetable}{cccc}
\tablecaption{Data of \citet{junker98}.\label{tab:jun}} 
\tablewidth{0pt}
\tablehead{
$E_{c.m.}$   & $S$  &  ${\Delta}S_\mathrm{stat}$  &  ${\Delta}S_\mathrm{sys}$  \\
(MeV)  & (MeVb) &   (MeVb) &   (MeVb) 
}
\startdata
0.02076 & 6.80 & 0.82 & 0.28 \\
0.02123 & 7.15 & 1.06 & 0.29 \\
0.02175 & 7.63 & 0.91 & 0.31 \\
0.02228 & 5.85 & 0.89 & 0.24 \\
0.02233 & 7.27 & 1.05 & 0.40 \\
0.02278 & 5.97 & 0.64 & 0.24 \\
0.02282 & 7.21 & 0.84 & 0.39 \\ 
0.02315 & 6.82 & 1.47 & 0.42 \\
0.02321 & 7.50 & 1.02 & 0.30 \\
0.02370 & 6.87 & 0.74 & 0.26 \\
0.02425 & 6.66 & 0.74 & 0.26 \\
0.02430 & 6.90 & 0.72 & 0.37 \\
0.02452 & 7.10 & 0.79 & 0.31 \\
0.02470 & 6.23 & 0.37 & 0.24 \\
0.02480 & 5.96 & 0.62 & 0.23 \\
0.04582 & 6.14 & 0.23 & 0.39 \\
0.05064 & 5.63 & 0.14 & 0.31 \\
0.05594 & 5.50 & 0.16 & 0.29 \\
0.06106 & 5.41 & 0.14 & 0.26 \\
0.06606 & 5.43 & 0.15 & 0.26 \\
0.07122 & 5.43 & 0.14 & 0.26 \\
0.07629 & 5.32 & 0.11 & 0.22 \\
0.08150 & 5.33 & 0.12 & 0.22 \\
0.08651 & 5.23 & 0.11 & 0.22 \\
0.09170 & 5.15 & 0.11 & 0.21 \\
\enddata
\end{deluxetable}

\begin{deluxetable}{cc|cc}
\tablecaption{Data of \citet{krauss87}.\label{tab:kra}} 
\tablewidth{0pt}
\tablehead{
$E_{c.m.}$   & $S \pm \Delta S_{\mathrm{stat}}\tablenotemark{a}$  & $E_{c.m.}$   & $S \pm \Delta S_{\mathrm{stat}}\tablenotemark{a}$  \\
(MeV) & (MeVb) & (MeV) & (MeVb)
}
\startdata
0.02451   &  5.07  $\pm$   1.34 &0.0863     &  4.92  $\pm$   0.13 \\
0.02655   &  5.18  $\pm$   1.06  &0.0895     &  5.31  $\pm$   0.30 \\
0.02900   &  5.23  $\pm$   0.58  &0.0916     &  4.69  $\pm$   0.07 \\
0.03145   &  5.45  $\pm$   0.45 &0.0940     &  4.86  $\pm$   0.08 \\
0.03390   &  5.26  $\pm$   0.52 &0.0972     &  4.97  $\pm$   0.08 \\
0.03634   &  5.35  $\pm$   0.41 &0.1034     &  4.93  $\pm$   0.10 \\
0.03909   &  5.77  $\pm$   0.35 &0.1092     &   4.77 $\pm$   0.16 \\
0.04124   &  5.03  $\pm$   0.43 &0.1160     &  4.89  $\pm$    0.08 \\
0.04373   &  4.88  $\pm$   0.24 &0.1215     &  4.67  $\pm$    0.08 \\
0.04648   &  4.98  $\pm$   0.26 &0.1336     &  4.56  $\pm$    0.13 \\
0.04808   &  5.08  $\pm$   0.16 & 0.1413     &  4.62  $\pm$    0.09 \\
0.04900   &  5.06  $\pm$   0.19 &0.1460     &  4.97  $\pm$    0.10 \\
0.04932   &  5.86  $\pm$   0.32 &0.1563     &  4.63  $\pm$    0.05 \\
0.0544     &  5.71  $\pm$   0.32 &0.1579     &  4.56  $\pm$    0.08 \\
0.0594     &  5.10  $\pm$   0.36 &0.1689     &  4.67  $\pm$    0.05 \\
0.0644     &  5.18  $\pm$   0.20 &0.1705     &  4.73  $\pm$    0.05 \\ 
0.0646     &  5.56  $\pm$   0.23 &0.1954     &  4.68  $\pm$    0.21 \\
0.0680     &  5.39  $\pm$   0.31 &0.2198     &  4.35  $\pm$    0.22 \\
0.0693     &  5.93  $\pm$   0.17 &0.2443     &  4.57  $\pm$    0.19 \\
0.0727     &  5.30  $\pm$   0.18 &0.2688     &  4.73  $\pm$    0.26 \\
0.0734     &  5.55  $\pm$   0.25 &0.2933     &  5.09  $\pm$    0.28 \\
0.0778     &  5.27  $\pm$   0.20 &0.3179     &  4.40  $\pm$    0.26 \\
0.0794     &  5.26  $\pm$   0.18 &0.3425     &  4.41  $\pm$    0.24 \\
0.0845     &  5.12  $\pm$   0.18 & &\\
\enddata
\tablenotetext{a}{Systematic uncertainty: 4.5\%.}
\end{deluxetable}

\subsubsection{Data of \citet[][]{dwarakanath71}}
We included the data of \citet[][]{dwarakanath71} in our analysis. Statistical and systematic uncertainties are not reported directly in that work, but it is possible to estimate the respective contributions from the information provided. Experimental S-factors versus center-of-mass energy are shown in their Figure 8 and we extracted the $17$ data points directly from the figure. Three data points with error bars are shown in representative energy regions. Their energies and S-factors are $S(0.126~\mathrm{MeV})$ $=$ $4.88$ ($\pm$ $10.0$\%), $S(0.489~\mathrm{MeV})$ $=$ $4.13$ ($\pm$ $8.0$\%), and $S(0.997~\mathrm{MeV})$ $=$ $3.67$ ($\pm$ $8.2$\%). The error bars ``include statistical and estimated systematic errors in both measured total cross sections and center-of-mass energy''. 

The {\it total} cross section in \citet[][]{dwarakanath71} is approximately given by 4$\pi$ times the differential cross section at $90^\circ$. Their Figure~6 shows the differential cross section at a center-of-mass energy of $150$~keV. Representative uncertainties are shown at two laboratory angles ($50^\circ$ and $130^\circ$). Each of these consists of two error bars, ``the larger error bar indicates the absolute error in the measured differential cross section and the smaller error bar indicates the relative error between measurements at different angles''. Extracting these values from their figure, we find for the relative (i.e., statistical) uncertainty a value of 5.5\% and for the absolute (statistical and systematic) uncertainty a value of 9.9\%. Assuming that statistical and systematic uncertainties have been added quadratically in \citet[][]{dwarakanath71}, we find a systematic uncertainty of 8.2\% at a center-of-mass energy near 150 keV. Therefore, we adopt a global systematic uncertainty of 8.2\% for all $17$ data points. For the statistical uncertainty we assume a value of 4.0\% at energies above $300$~keV, and $7.0$\% at lower energies. These estimates agree with the overall uncertainties quoted above for the three total S-factors. Our adopted values are listed in Table~\ref{tab:he3he3}.

\begin{deluxetable}{cccc}
%\tabletypesize{\scriptsize}
%\rotate
\tablecaption{Our adopted data from \citet[][]{dwarakanath71}.\label{tab:he3he3}}
\tablewidth{0pt}
\tablehead{
 $E_{c.m.}$\tablenotemark{a}  &   $S$\tablenotemark{a}   &   $\Delta S_{\mathrm{stat}}$\tablenotemark{b}   & $\Delta S_{\mathrm{sys}}$\tablenotemark{b}  \\
          (MeV)  &   (MeV~b)   &  (\%)    & (\%)  
%\cline{2-3} \cline{5-6}
}
\startdata
 0.088 	&  4.86  &  7.0  &  8.2   	 \\
 0.126	&  4.88  &  7.0  &  8.2   	 \\
 0.155	&  4.96  &  7.0  &  8.2   	 \\
 0.193	&  4.51  &  7.0  &  8.2   	 \\
 0.234	&  4.68  &  7.0  &  8.2   	 \\
 0.288	&  4.45  &  7.0  &  8.2   	 \\
 0.338	&  4.34  &  4.0  &  8.2   	 \\
 0.379	&  4.50  &  4.0  &  8.2   	 \\
 0.435	&  4.21  &  4.0  &  8.2    	\\
 0.488	&  4.13  &  4.0  &  8.2   	 \\
 0.591	&  3.90  &  4.0  &  8.2   	 \\
 0.691	&  3.76  &  4.0  &  8.2   	 \\
 0.746	&  3.70  &  4.0  &  8.2   	 \\
 0.792	&  3.50  &  4.0  &  8.2   	 \\
 0.895	&  3.51  &  4.0  &  8.2   	 \\
 0.997	&  3.66  &  4.0  &  8.2   	 \\
 1.081	&  3.50  &  4.0  &  8.2   	 \\
\enddata
%% Text for table notes should follow after the \enddata but before
%% the \end{deluxetable}. Make sure there is at least one \tablenotemark
%% in the table for each \tablenotetext.
%\tablecomments{Listed values are adopted from a galactic chemical evolution model that reproduces observed abundances in field stars of the same metallicity as NGC 2419 ([Fe/H] = $-2.1$).}
\tablenotetext{a}{Extracted from Figure~8 of \citet[][]{dwarakanath71}.} 
\tablenotetext{b}{See discussion in Appendix~\ref{app:he3he3}.}
\end{deluxetable}

\subsubsection{Other data}
Several data sets that were used in previous evaluations \citep[][]{angulo99,descouvemont04,adelberger11,xu13} were not incorporated into our analysis. For example, \citet{brown87} measured the cross section at energies too high to be of interest here, while \citet{Dwa74,Bac67,Wan66} did not provide enough information to reliably estimate the separate contributions of statistical and systematic uncertainties. 

\subsection{The $^3$He($\alpha$,$\gamma$)$^7$Be reaction}\label{app:3heag}
\subsubsection{The data of \citet[][]{brown07}}
We extracted only the activation data of \citet{brown07} from their Table III, and list the values here in Table~\ref{tab:bro}. A systematic uncertainty of $3.0$\% is adopted from their Table~IV.

\begin{deluxetable}{cc}
\tablecaption{Activation data of \cite{brown07}.\label{tab:bro}} 
\tablewidth{0pt}
\tablehead{
$E_{c.m.}$  &  $S \pm \Delta S_{\mathrm{stat}}\tablenotemark{a}$    \\
 (MeV)  &   (keVb)  
}
\startdata
0.3274  $\pm$   0.0013  & 0.495  $\pm$   0.015 \\
0.4260  $\pm$   0.0004 & 0.458  $\pm$   0.010 \\
0.5180  $\pm$   0.0005 & 0.440   $\pm$  0.010 \\
0.5815  $\pm$   0.0008 & 0.400  $\pm$   0.011  \\
0.7024  $\pm$   0.0006 & 0.375  $\pm$   0.010  \\
0.7968  $\pm$   0.0003 & 0.363  $\pm$   0.007 \\
1.2337  $\pm$   0.0003 & 0.330  $\pm$   0.006  \\
1.2347  $\pm$   0.0003 & 0.324   $\pm$  0.006  \\
\enddata
\tablenotetext{a}{Systematic uncertainty: 3.0\%.}
\end{deluxetable}

\subsubsection{Data of \citet[][]{narasingh04}}
We use the four activation data points of \citet{narasingh04}; see their Fig.~3 and Table~II. The statistical and systematic uncertainties are taken from their Table~II. The adopted results are listed in our Table~\ref{tab:nar}.

\begin{deluxetable}{cccc}
\tablecaption{Data of  \citet{narasingh04}.\label{tab:nar}} 
\tablewidth{0pt}
\tablehead{
$E_{c.m.}$ & $S$  &  ${\Delta}S_\mathrm{stat}$  &  ${\Delta}S_\mathrm{sys}$ \\
 (keV)  &  (keVb) &   (keVb) &   (keVb)
}
\startdata
      0.420   &    0.420   &   0.014   &  0.030  \\      
       0.506   &    0.379  &   0.015    &   0.027  \\
       0.615   &   0.362   &   0.010    &   0.015\\
        0.950 & 0.316   &   0.006    &   0.007 \\
\enddata
\end{deluxetable}

\subsubsection{Data of \citet[][]{dileva09}}
The cross section data are taken from Table I of \citet{dileva09}. The corresponding S-factor values, computed using Eqs.~\ref{eq:sig2s1} and \ref{eq:sig2s2}, are displayed in Table~\ref{tab:dil}, together with statistical uncertainties. The total systematic uncertainty of 5\% is dominated by contributions from the target thickness (4\%) and the current integration (1\%).

\begin{deluxetable}{ccc}
\tablecaption{Recoil data of \citet{dileva09}.\label{tab:dil}} 
\tablewidth{0pt}
\tablehead{
$E_{CM}$   &  $\sigma$   & $S \pm \Delta S_{\mathrm{stat}}\tablenotemark{a}$ \\
(MeV)  &  ($\mu$b)  & (keVb)
}
\startdata
   0.701 &  1.140   $\pm$  0.200 & 0.393   $\pm$  0.069 \\
   0.802  & 1.460   $\pm$  0.080  & 0.385  $\pm$   0.021\\
   0.902  & 1.590   $\pm$  0.070 &  0.339  $\pm$   0.015 \\
   1.002  & 1.960  $\pm$   0.070 &  0.351  $\pm$   0.013 \\
   1.002  & 1.860  $\pm$   0.060 &  0.333  $\pm$   0.011\\
   1.102 &  2.160   $\pm$  0.020 &  0.334  $\pm$   0.003\\
   1.102  & 2.190  $\pm$   0.040 &  0.339  $\pm$   0.006\\
   1.103  & 2.160  $\pm$   0.060 &  0.334  $\pm$   0.009\\
   1.203 &  2.440  $\pm$   0.050 &  0.333   $\pm$  0.007\\
   1.203 &  2.440  $\pm$   0.090 &  0.333   $\pm$  0.012\\
   1.353  & 2.790   $\pm$  0.070 &  0.327  $\pm$   0.008\\
   1.403 &  3.060  $\pm$   0.040  & 0.343  $\pm$   0.004\\
   1.403 &  3.030   $\pm$  0.080 &  0.340   $\pm$  0.009\\
   1.403 &  3.060  $\pm$    0.100 &  0.343   $\pm$  0.011\\
   1.504  & 3.270   $\pm$  0.100  & 0.339  $\pm$   0.010\\
\enddata
\tablenotetext{a}{Systematic uncertainty: 5.0\%.}
\end{deluxetable}

\subsubsection{The LUNA data}
The results of activation measurements at LUNA are presented in \citet{bemmerer06} and \citet{gyurky07}. They are summarized in Table~2 of \citet{costantini08} and are listed in our Table~\ref{tab:luna}.

\begin{deluxetable}{ccll}
\tablecaption{Activation data of \citet{costantini08}.\label{tab:luna}} 
\tablewidth{0pt}
\tablehead{
$E_{c.m.}$ & $S$  &  ${\Delta}S_\mathrm{stat}$  &  ${\Delta}S_\mathrm{sys}$ \\
 (MeV)  &  (keVb) &   (keVb) &   (keVb)
}
\startdata
 0.0929  & 0.534 & 0.016 & 0.017 \\
0.1057 & 0.493 & 0.015 & 0.015 \\
 0.1265   &    0.514   & 0.020 &     0.030 \\ 
 0.1477    &   0.499    & 0.017 & 0.030 \\
  0.1689    &   0.482 & 0.020 & 0.030 \\
  0.1695    &   0.507 & 0.010 & 0.015 \\
\enddata
\end{deluxetable}

\bibliographystyle{aasjournal}

\end{document}